\documentclass[final,5p,times,twocolumn]{elsarticle}

\usepackage{amssymb}
\usepackage{amsmath}
\usepackage{subcaption}
\usepackage{float} 
\usepackage{url}
\usepackage{adjustbox}
\usepackage{hyperref}
\usepackage{ulem}

\hypersetup{
    colorlinks=true,
    linkcolor=blue,
    citecolor=blue,
    urlcolor=blue
}

\makeatletter
\def\ps@pprintTitle{%
    \let\@oddfoot\@empty
    \let\@evenfoot\@empty
    \def\@oddhead{%
        \hfil
        \footnotesize\itshape
        Prepared for submission to \textit{Physics of the Dark Universe}
        \quad -- \quad
        DOI: 
        {XXXXXXXXXXXXXXXXX}
        \hfil
    }%
    \let\@evenhead\@oddhead
}
\makeatother

\journal{Physics of the Dark Universe}

\begin{document}

\begin{frontmatter}


\title{Cosmography with gravitational waves II: Padé approximants in an extended redshift regime}

\author[label1]{Jonathan Morais}
\ead{jmorais@on.br}

\author[label2,label1]{Rodrigo Gon\c{c}alves}
\ead{rsg\_goncalves@ufrrj.br}

\author[label1]{Jailson Alcaniz}
\ead{alcaniz@on.br}

\affiliation[label1]{organization={Observatório Nacional},
            city={Rio de Janeiro},
            postcode={20921-400},
            state={RJ},
            country={Brasil}}

\affiliation[label2]{organization={Departamento de Física, Universidade Federal Rural do Rio de Janeiro},
            city={Seropédica},
            postcode={23897-000},
            state={RJ},
            country={Brasil}}
            
\begin{abstract}
We evaluate the capability of next-generation gravitational-wave detectors, including LISA, the Einstein Telescope, and DECIGO, to probe cosmographic parameters using electromagnetically bright standard sirens. Adopting Padé approximants, we extend the cosmographic analysis to substantially higher redshifts and constrain the parameters $H_0$, $q_0$, $j_0$, $s_0$, and $l_0$. We examine the constraints primarily as a function of the number of detections and, in a complementary analysis, across correlated SNR--redshift regimes. LISA provides percent-level constraints on $H_0$, while the Einstein Telescope and DECIGO reach subpercent precision as the catalogue size increases. The parameter $q_0$ can be constrained at the tens-of-percent level, whereas the higher-order parameters $j_0$, $s_0$, and $l_0$ remain progressively less determined. Compared with the low-redshift Taylor analysis, the Padé approach extends the usable redshift range, increases the accessible event population and enables a higher-order reconstruction of the expansion history.
\end{abstract}


\end{frontmatter}




\section{Introduction}

\indent Gravitational-wave (GW) astronomy has established a new observational window for investigating the Universe, with applications ranging from compact-object astrophysics and tests of gravity to cosmology~\cite{2017Abbott}. Since the first direct detection of a GW signal by the Laser Interferometer Gravitational-Wave Observatory (LIGO) in 2015~\cite{LIGOScientific:2016aoc}, the network of second-generation detectors, including Advanced LIGO (aLIGO) ~\cite{LIGOScientific:2014pky}, Virgo~\cite{VIRGO:2014yos}, and KAGRA~\cite{Somiya:2011np}, has reported hundreds of compact-binary coalescences~\cite{LIGOScientific:2025snk}. The next generation of GW observatories is expected to substantially increase both the number of detections and the redshift range accessible to gravitational-wave cosmology.\\
\indent Among these future facilities, the Einstein Telescope (ET)~\cite{Punturo:2010zz} will extend ground-based observations of stellar-origin compact binaries to cosmological distances, while the space-based Deci-hertz Interferometer Gravitational-wave Observatory (DECIGO)~\cite{Kawamura:2006up} will probe the decihertz frequency band with high sensitivity. At lower frequencies, the Laser Interferometer Space Antenna (LISA) will open the millihertz gravitational-wave window and detect, among other systems, massive black-hole binaries over a broad cosmological redshift range~\cite{LISA:2024hlh}. Together, these observatories will provide access to source populations and redshift regimes extending considerably beyond those currently available to gravitational-wave cosmology.\\
\indent Compact-binary coalescences are particularly relevant in this context because they can be used as \textit{standard sirens}~\cite{Schutz:1986gp}. The amplitude and phase evolution of the GW signal allow the luminosity distance to be inferred directly, without relying on the cosmic distance ladder. The source redshift, however, cannot in general be determined from the gravitational-wave signal alone because of its degeneracy with the intrinsic binary parameters. When an electromagnetic counterpart is identified, the host galaxy can be associated with the GW source and an independent redshift measurement can be obtained. The binary neutron-star merger GW170817, observed together with the short gamma-ray burst GRB170817A~\cite{LIGOScientific:2017zic}, provided the first observational realization of this scenario.\\
\indent Events for which both the luminosity distance and the redshift are available, usually referred to as \textit{bright sirens}~\cite{Holz:2005df}, provide a direct mapping of the luminosity-distance--redshift relation $d_L(z)$. Future ground- and space-based detectors are expected to substantially enlarge the samples of GW sources for which cosmological information can be extracted~\cite{Sathyaprakash:2009xs,Zhang:2018byx,Wang:2021srv,Yang:2021qge}. For LISA, massive black-hole binaries constitute a particularly relevant population for cosmology, and possible electromagnetic counterparts could enable a subset of these events to be used as bright standard sirens~\cite{Mangiagli:2022niy}. The increasing redshift reach of future observatories therefore motivates cosmological descriptions capable of remaining reliable well beyond the nearby Universe.\\
\indent In this context, cosmography provides a natural framework for describing the luminosity-distance--redshift relation without assuming a specific dynamical model for the cosmic expansion~\cite{Weinberg:1972kfs}. Rather than specifying the matter content or gravitational dynamics of the Universe, cosmography characterizes its expansion kinematically through quantities associated with successive time derivatives of the scale factor. In its standard formulation, cosmological observables are expressed as Taylor series in the redshift~\cite{Visser:2004bf}. Although this approach is particularly useful at low redshift, its applicability becomes increasingly limited as more distant sources are included.\\
\indent In a previous communication, hereafter Paper I~\cite{Morais:2026qdw}, we investigated the capability of bright standard sirens observed by aLIGO, ET and DECIGO to constrain the Hubble constant, the deceleration and jerk parameters using the standard Taylor cosmographic expansion. We adopted a third-order expansion and quantified its range of validity by requiring the truncated luminosity distance to differ by less than $1\%$ from the corresponding fiducial $\Lambda$CDM prediction. Under this criterion, the analysis was restricted to approximately $z_{\max}\simeq0.7$. This conservative choice ensured that the truncation error remained under control, but necessarily excluded a substantial fraction of the sources potentially accessible to future detectors.\\
\indent The consequences of this redshift restriction depend strongly on the observational reach of each interferometer. For aLIGO, whose detectable bright-siren population is already confined to approximately $z\lesssim0.2$~\cite{Lagos:2019kds}, the $z_{\max}\simeq0.7$ limit imposed by the Taylor expansion does not restrict the accessible source population. Extending the validity of the cosmographic parametrization to higher redshifts is therefore not expected to provide a significant observational gain for it. For this reason, and because the main purpose of the present work is precisely to investigate the additional cosmological information made accessible by extending the redshift range, we do not include Advanced LIGO in the analyses performed in this second paper.\\
\indent The situation is markedly different for future observatories. Both ET and DECIGO are expected to detect large populations of compact binaries beyond the low-redshift domain explored in Paper I, so that the Taylor cutoff excludes a significant fraction of their potentially useful cosmological events. This limitation is even more consequential for LISA. While LISA is designed to observe massive black-hole binaries over a broad cosmological redshift interval, only a very small fraction of the relevant source population lies within the $z\lesssim0.7$ domain adopted in Paper I. Consequently, LISA was not included in that analysis. The limited redshift validity of the Taylor expansion therefore affected not only the number of usable events, but also the range of GW observatories and source populations that could be meaningfully investigated within the same cosmographic framework.\\
\indent The origin of this limitation lies in the convergence properties of the standard Taylor expansion in redshift. Since the expansion is performed around $z=0$, its accuracy progressively deteriorates as sources at larger redshift are included, and the conventional series becomes increasingly unreliable at redshifts of order unity and beyond~\cite{Visser:2004bf,deSouza:2021xtg}. Increasing the truncation order can reduce the approximation error over part of this interval, but each additional term introduces a higher-order cosmographic parameter and consequently increases the dimensionality of the parameter space and the associated degeneracies. Fully exploiting the redshift reach expected for ET, DECIGO, and particularly LISA therefore requires an alternative representation of the luminosity-distance relation.\\
\indent In this second paper, we address this limitation by adopting Padé approximants~\cite{Gruber:2013wua}. Instead of representing the luminosity distance by a finite polynomial, Padé cosmography describes it through ratios of polynomials, which can provide a more accurate representation of the underlying expansion history over a wider redshift interval. Previous studies have shown that suitable Padé parametrizations can improve the behavior of cosmographic observables at intermediate and high redshifts and mitigate the limitations associated with the standard Taylor expansion~\cite{Capozziello:2020ctn,Hu:2022udt}. Nevertheless, a Padé parametrization does not guarantee accuracy over an arbitrary redshift range, and the domain of validity of each approximant must be explicitly quantified before it is used for parameter inference.\\
\indent The primary goal of this work is therefore to determine how far in redshift the cosmographic analysis of Paper I can be extended when the Taylor description is replaced by Padé approximants. We first quantify the accuracy and range of validity of different Padé parametrizations relative to the same fiducial $\Lambda$CDM cosmology adopted previously. We then investigate the cosmographic constraints obtainable from simulated bright-siren catalogues for ET and DECIGO, whose accessible event populations extend significantly beyond the Taylor regime considered in Paper I. Most importantly, we include LISA for the first time in this series, allowing us to explore a complementary population of massive black-hole binaries and a substantially broader cosmological redshift interval. In this way, we assess whether the improved high-redshift behavior of Padé cosmography allows gravitational-wave information that was inaccessible within the low-redshift Taylor framework of Paper I to be incorporated into cosmographic parameter estimation.

\section{Cosmography with Padé Approximants}

\indent Cosmography provides a purely kinematical description of the cosmic expansion, relying on the assumptions of homogeneity and isotropy without specifying the underlying gravitational dynamics or matter content. As in Paper I~\cite{Morais:2026qdw}, we assume a spatially flat Friedmann--Lemaître--Robertson--Walker (FLRW) geometry and characterize the expansion history through successive time derivatives of the scale factor evaluated at the present cosmic time. For completeness, and to establish the notation adopted throughout this work, we briefly summarize the cosmographic framework before introducing its Padé representation.\\
\indent Following~\cite{Cattoen:2007sk}, the scale factor can be expanded around the present cosmic time $t_0$ as
\begin{align} \label{cosmography_a(t)}
\frac{a(t)}{a(t_0)} &={} 1 + H_0 (t - t_0) - \frac{1}{2} q_0 [H_0 (t - t_0)]^2 
+ \frac{1}{6} j_0 [H_0 (t - t_0)]^3 \nonumber \\
\quad &+ \frac{1}{24} s_0 [H_0 (t - t_0)]^4 
+ \frac{1}{120} l_0 [H_0 (t - t_0)]^5  + \mathcal{O}(6),
\end{align}
where $H_0$ is the Hubble constant at the present epoch, while the higher-order derivatives of the scale factor define the cosmographic parameters
\begin{align} \label{cosmographic_parameters}
q_0 &= -\frac{1}{H_0^2} \left( \frac{1}{a} \frac{d^2 a}{dt^2} \right)\Big|_{t=t_0}, 
& j_0 &= \frac{1}{H_0^3} \left( \frac{1}{a} \frac{d^3 a}{dt^3} \right)\Big|_{t=t_0}, \nonumber \\
s_0 &= \frac{1}{H_0^4} \left( \frac{1}{a} \frac{d^4 a}{dt^4} \right)\Big|_{t=t_0}, 
& l_0 &= \frac{1}{H_0^5} \left( \frac{1}{a} \frac{d^5 a}{dt^5} \right)\Big|_{t=t_0}.
\end{align}
The parameters $q_0$, $j_0$, $s_0$, and $l_0$ are respectively referred to as the \textit{deceleration}, \textit{jerk}, \textit{snap}, and \textit{lerk} parameters. Since they are defined exclusively through derivatives of the scale factor, these quantities provide a model-independent kinematical characterization of the expansion history around the present epoch.\\
\indent Using the definition of redshift, $1+z \equiv a(t_0)/a(t)$, the cosmographic expansion can be rewritten in terms of directly observable quantities. In particular, the Hubble parameter as a function of redshift is given by~\cite{Zhang:2016urt}
\begin{align}
\label{H(z)}
H(z) &= H_0 \Bigg[\,
1 + (1+q_0)\,z
+ \frac{1}{2}\left(j_0-q_0^2\right)z^2 +
\nonumber \\[4pt]
&\quad
+ \frac{1}{6}\Bigl(
3q_0^2 + 3q_0^3
-4q_0j_0 -3j_0 -s_0
\Bigr)z^3 + 
\nonumber \\[4pt]
&\quad
+ \frac{1}{24}\Bigl(
-12q_0^2 -24q_0^3 -15q_0^4
+32q_0j_0 +25q_0^2j_0 +
\nonumber \\[4pt]
&\qquad
+7q_0s_0 +12j_0 -4j_0^2
+8s_0 +l_0
\Bigr)z^4 +
\nonumber \\[4pt]
&\quad
+ \frac{1}{120}\Bigl(
m_0
-60q_0^2
-180q_0^3
-225q_0^4
-105q_0^5 +
\nonumber \\[4pt]
&\qquad
-10j_0^2(6+7q_0)
+l_0(15+11q_0) +
\nonumber \\[4pt]
&\qquad
+15j_0\bigl(
4+16q_0+25q_0^2+14q_0^3-s_0
\bigr) +
\nonumber \\[4pt]
&\qquad
+60s_0
+105q_0s_0
+60q_0^2s_0
\Bigr)z^5
+ \mathcal{O}(z^6)
\,\Bigg].
\end{align}
The same procedure can be applied to other cosmological observables, in particular to the luminosity distance $d_L(z)$, which constitutes the main observable in the standard-siren analysis performed in this work.\\
\indent Despite its conceptual simplicity, the conventional cosmographic approach based on Taylor expansions in redshift suffers from intrinsic convergence limitations. As discussed in~\cite{Cattoen:2007sk}, the Taylor expansion of cosmological observables around $z=0$ has a finite radius of convergence. In particular, the relation $1+z=a(t_0)/a(t)$ introduces the special point $z=-1$, associated with the limit of an infinitely expanded Universe, implying that the standard Taylor representation cannot in general be expected to converge beyond $|z|=1$. Consequently, although a truncated expansion can formally be evaluated at larger redshifts, its accuracy progressively deteriorates as increasingly distant sources are included, and truncation effects become particularly relevant at redshifts of order unity and beyond.\\
\indent Several alternative parametrizations have been proposed to alleviate the limitations of the standard redshift expansion~\cite{Gruber:2013wua,Cattoen:2007sk,Aviles:2012ay,Capozziello:2017nbu}. Among them, rational approximations have attracted particular attention because of their improved behaviour at intermediate and high redshifts~\cite{Capozziello:2020ctn,Hu:2022udt}. In particular, Padé approximants are constructed by matching the coefficients of a Taylor expansion while replacing the finite polynomial representation by a ratio of two polynomials. Their rational structure can provide an accurate representation of the underlying function over a wider interval than the corresponding truncated Taylor polynomial and may remain well behaved in regions where the original power-series representation becomes inadequate. This property, however, is not guaranteed for an arbitrary Padé order, and each parametrization must be examined individually before being applied to cosmological data.\\
\indent More explicitly, consider the Taylor expansion of a generic function around $z=0$,
\begin{equation}
f(z)=\sum_{i=0}^{\infty}c_i z^i,
\end{equation}
where $c_i=f^{(i)}(0)/i!$. A Padé approximant of order $(n,m)$ is defined as the rational function~\cite{Capozziello:2020ctn}
\begin{equation} \label{pade_main_eq}
P_{n,m}(z)=\frac{\displaystyle\sum_{i=0}^{n}a_i z^i}{\displaystyle 1+\sum_{j=1}^{m}b_j z^j}.
\end{equation}
Here, $n$ and $m$ denote the degrees of the numerator and denominator polynomials, respectively, and are non-negative integers. The constant term of the denominator is conventionally normalized to unity, $b_0=1$, removing the overall multiplicative degeneracy between numerator and denominator. The rational function therefore contains $n+m+1$ coefficients before the matching conditions are imposed.\\
\indent The Padé coefficients are determined by requiring the expansion of $P_{n,m}(z)$ around $z=0$ to reproduce the original Taylor series through order $n+m$, such that
\begin{equation}
f(z)-P_{n,m}(z)=\mathcal{O}\left(z^{n+m+1}\right).
\end{equation}
Equivalently, multiplying Eq.~\eqref{pade_main_eq} by its denominator and matching equal powers of $z$ yields the following system for the denominator coefficients~\cite{Rocha:2022gog}:
\begin{equation}
\sum_{j=1}^{m}b_jc_{n+k-j}=-c_{n+k}, \qquad k=1,\ldots,m,
\end{equation}
Once the coefficients $b_j$ are determined, the numerator coefficients follow from
\begin{equation}
a_i=\sum_{j=0}^{\min(i,m)}b_jc_{i-j}, \qquad i=0,\ldots,n.
\end{equation}
Thus, the coefficients appearing in the Padé representation are completely determined by the coefficients of the corresponding Taylor expansion and do not constitute additional independent cosmological parameters.\\
\indent We now apply this formalism to the luminosity distance $d_L(z)$. The cosmographic Taylor expansion derived in Paper I~\cite{Morais:2026qdw} is used as the generating series from which the different Padé approximants are constructed, following the general approach adopted in previous applications of Padé cosmography to the luminosity distance~\cite{Aviles:2014pade,Zhou:2016pade}. For a given $P_{n,m}$ parametrization, only the Taylor coefficients up to order $n+m$ are required. Since these coefficients are themselves functions of the cosmographic quantities, the resulting coefficients $a_i$ and $b_j$ can be expressed entirely in terms of $H_0$ and the successive cosmographic parameters required at the corresponding order.\\
\indent An important consequence is that different combinations of $(n,m)$ generally depend on different sets of cosmographic parameters. Lower-order approximants involve only the lowest derivatives of the scale factor, whereas higher-order combinations progressively introduce parameters such as $s_0$, $l_0$, and, when required, still higher-order quantities. Increasing the order of the rational approximation may therefore improve the representation of $d_L(z)$ over a wider redshift interval, but it simultaneously enlarges the cosmographic parameter space and can increase the statistical uncertainties and degeneracies among the inferred quantities~\cite{Aviles:2014pade,Zhou:2016pade}.\\
\indent Moreover, the behaviour of a Padé approximant is determined not only by its total order $n+m$, but also by how that order is distributed between the numerator and denominator. Different combinations may therefore provide substantially different approximations even when they are constructed from the same number of Taylor coefficients~\cite{Capozziello:2020ctn,Petreca:2024pade}. In particular, zeros of the denominator can generate poles within the redshift interval of interest, while other combinations may display an unsuitable evolution of the luminosity distance or fail to improve significantly upon the corresponding Taylor representation~\cite{Zhou:2016pade,Petreca:2024pade}. The choice of a Padé parametrization must consequently be based on its actual cosmological behaviour rather than solely on its formal order.\\
\indent For this reason, we do not select a specific Padé combination a priori. Following the strategy adopted in systematic studies of different Padé orders~\cite{Capozziello:2020ctn,Petreca:2024pade}, we consider several combinations of the numerator and denominator orders $(n,m)$ and assess their performance before deciding which parametrizations are suitable for the gravitational-wave analysis. Similar comparisons among different Padé orders have also been performed using high-redshift cosmological observables, showing that the performance of the approximation can depend significantly on the adopted order~\cite{Pourojaghi:2022pade}.\\
\indent The next step is therefore to determine the effective domain of validity of the different Padé parametrizations. In the following section, we systematically compare their luminosity-distance predictions with the same fiducial $\Lambda$CDM cosmology adopted in Paper I, quantify the relative deviation as a function of redshift, and examine the possible occurrence of pathological behaviour. Comparisons between Padé cosmography and fiducial cosmological predictions over extended redshift intervals have been used as a direct test of the accuracy of the approximation in previous studies~\cite{Aviles:2014pade,Pourojaghi:2022pade}. This procedure allows us to identify which Padé approximants provide an adequate description of $d_L(z)$ and to determine the maximum redshift that can be consistently adopted for each case. Only after this analysis are the selected parametrizations used to construct the simulated bright-siren catalogues and perform the cosmographic parameter estimation.

\section{Methodology}
\label{sec:methodology}

\indent In order to quantify the effective redshift range of the Padé cosmographic description, we first define the fiducial cosmological model used as the reference throughout the analysis. We assume a spatially flat $\Lambda$CDM cosmology characterized by the matter density parameter $\Omega_m=0.311$ and the Hubble constant $H_0=67.66$ km\,s$^{-1}$\,Mpc$^{-1}$, consistent with the Planck 2018 results~\cite{Planck:2018vyg}. The corresponding theoretical luminosity distance is used as the reference against which the different Padé approximants are evaluated.\\
\indent The fiducial values of the cosmographic parameters are obtained from their analytical relations within a spatially flat $\Lambda$CDM background~\cite{Aviles:2012ay}. Substituting
\begin{equation}
    H(z)=H_0\sqrt{\Omega_m(1+z)^3+1-\Omega_m}
\end{equation}
into the cosmographic definitions gives
\begin{align}
q_0 &= \frac{3}{2}\Omega_m-1, \nonumber \\
j_0 &= 1, \nonumber \\
s_0 &= 1-\frac{9}{2}\Omega_m, \nonumber \\
l_0 &= 1+3\Omega_m+\frac{27}{2}\Omega_m^2 .
\label{LCDM_cosmography_pade}
\end{align}
For $\Omega_m=0.311$, the fiducial cosmographic parameters adopted in the analysis are summarized in Table~\ref{tab_fiducial_parameters_pade}.\\
\begin{table}[ht]
\centering
\begin{tabular}{ccccc}
\hline
$H_0$ & $q_0$ & $j_0$ & $s_0$ & $l_0$ \\
\hline
$67.66$ & $-0.533$ & $1$ & $-0.400$ & $3.239$ \\
\hline
\end{tabular}
\caption{Fiducial cosmological and cosmographic parameters adopted in the analysis. The Hubble constant is given in units of km\,s$^{-1}$\,Mpc$^{-1}$.}
\label{tab_fiducial_parameters_pade}
\end{table}

\indent As discussed in the previous section, different combinations of the numerator and denominator orders can lead to substantially different Padé representations of the luminosity distance. We therefore do not select a particular approximant solely on the basis of its formal order. Instead, the different combinations considered are first inspected over an extended redshift interval in order to identify pathological or physically unsuitable behaviours. In particular, we discard parametrizations that develop poles within the redshift range of interest, produce negative luminosity distances, or exhibit a non-monotonic high-redshift evolution in which $d_L(z)$ decreases as the redshift increases. Similar considerations regarding the occurrence of singularities and the physical behaviour of different Padé orders have been emphasized in previous cosmographic analyses~\cite{Capozziello:2020ctn, Aviles:2014pade}.\\
\begin{figure*}[!t]
    \centering

    \begin{subfigure}[t]{0.5\textwidth}
        \centering
        \includegraphics[width=\linewidth]{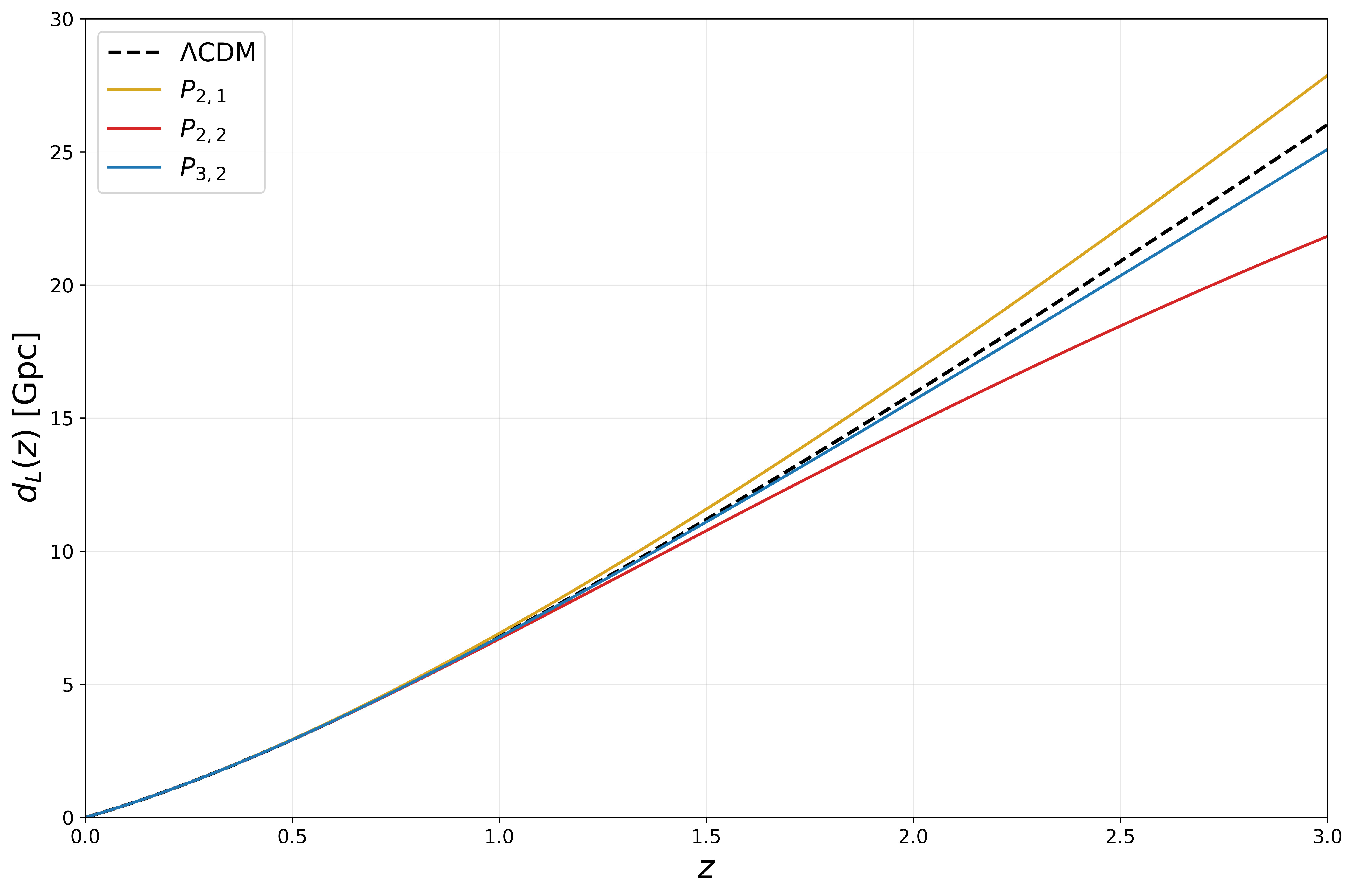}
        \label{fig:pade_dl_curves}
    \end{subfigure}
    \hfill
    \begin{subfigure}[t]{0.49\textwidth}
        \centering
        \includegraphics[width=\linewidth]{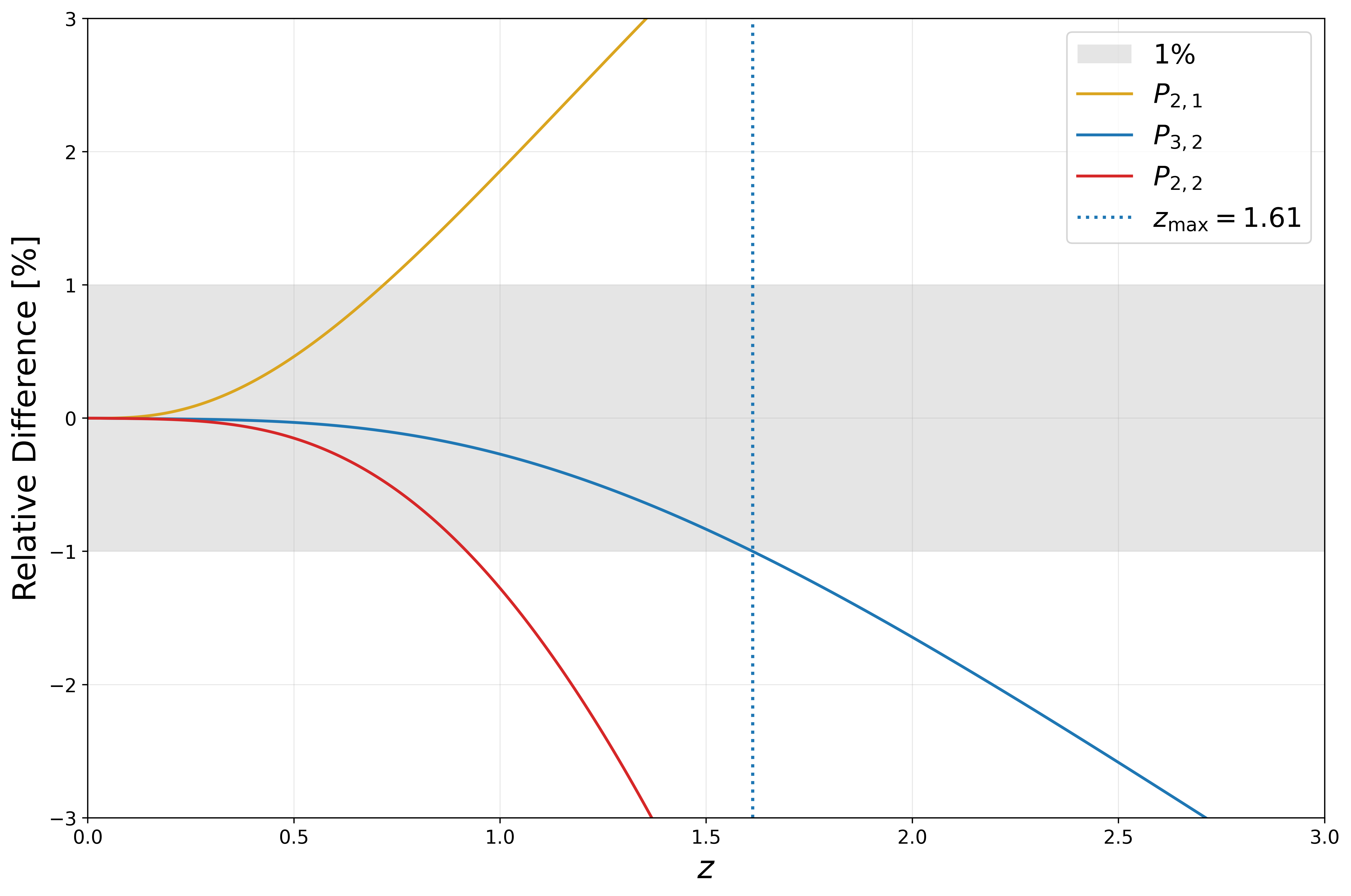}
        \label{fig:pade_dl_errors}
    \end{subfigure}

    \caption{Luminosity distance as a function of redshift for the $P_{2,1}$, $P_{2,2}$, and $P_{3,2}$ Padé approximants considered in the analysis (left panel), together with the fiducial $\Lambda$CDM prediction shown as a black dashed curve. The right panel shows the relative deviation of each Padé luminosity distance with respect to the fiducial model. The gray shaded region represents the adopted $1\%$ accuracy interval, while the vertical dotted line marks $z_{\max}\simeq1.61$, corresponding to the maximum redshift for which the $P_{3,2}$ approximation remains within this threshold.}
    \label{fig:pade_dl}
\end{figure*}
\indent After applying these requirements, three Padé approximants are retained for further comparison: $P_{2,1}$, $P_{2,2}$, and $P_{3,2}$. These parametrizations involve progressively higher orders of the cosmographic expansion. The $P_{2,1}$ approximant is constructed from the luminosity-distance series up to third order and consequently depends on $H_0$, $q_0$, and $j_0$, providing a direct counterpart to the third-order Taylor analysis performed in Paper I~\cite{Morais:2026qdw}. The $P_{2,2}$ approximant requires the expansion up to fourth order and therefore introduces the snap parameter $s_0$, whereas $P_{3,2}$ is constructed through fifth order and additionally depends on the lerk parameter $l_0$. Their explicit expressions are given in~\ref{app:pade_expressions}.\\
\indent The left panel of Fig.~\ref{fig:pade_dl} shows the luminosity distance predicted by these three Padé approximants over the extended interval $0<z<10$, together with the fiducial $\Lambda$CDM prediction. Although the three parametrizations remain free of poles and negative luminosity distances over the range displayed, their high-redshift behaviours differ substantially. In particular, $P_{2,2}$ exhibits a pronounced flattening at large $z$, departing qualitatively from the continued growth of the fiducial $\Lambda$CDM luminosity distance. This behaviour indicates that $P_{2,2}$ should not be regarded as a reliable approximation over the full redshift interval shown. Nevertheless, such a high-redshift departure does not preclude its use at lower redshifts, where the approximation may still reproduce the fiducial luminosity distance with sufficient accuracy. Therefore, the absence of pathological behaviour alone is not sufficient to determine the useful domain of each Padé approximant, and a quantitative accuracy criterion is required to establish the redshift interval over which each parametrization can be reliably employed.\\
\indent To quantify the agreement between each Padé representation and the fiducial cosmology, we define the relative deviation
\begin{equation}
    \Delta d_L(z)=100\times\left[
    \frac{d_L^{\text{Padé}}(z)-d_L^{\Lambda\mathrm{CDM}}(z)}
    {d_L^{\Lambda\mathrm{CDM}}(z)}
    \right].
\label{relative_deviation_pade}
\end{equation}
As in Paper I, we adopt a maximum absolute relative deviation of $1\%$ as the criterion defining the effective redshift domain of the cosmographic approximation. This choice provides a common accuracy threshold and allows a direct comparison between the redshift reach obtained with the Taylor and Padé descriptions.\\
\indent The relative deviations are shown in the right panel of Fig.~\ref{fig:pade_dl}. The $P_{2,1}$ approximant reaches the $1\%$ threshold at $z_{\max}\simeq0.72$, providing only a modest extension with respect to the third-order Taylor expansion considered in Paper I, whose corresponding limit was approximately $z_{\max}\simeq0.70$. The $P_{2,2}$ approximant remains within the adopted accuracy interval up to $z_{\max}\simeq0.92$, extending the useful range beyond the low-redshift Taylor regime. Among the physically well-behaved parametrizations considered here, however, $P_{3,2}$ provides the widest redshift interval satisfying the same criterion, reaching $z_{\max}\simeq1.61$.\\
\indent Since the main objective of this work is to investigate gravitational-wave cosmography over an extended redshift regime, we adopt $P_{3,2}$ as the fiducial Padé parametrization for the subsequent simulations and parameter-estimation analysis. It is worth mentioning that this choice should not be interpreted as establishing $P_{3,2}$ as universally preferable to other Padé orders, rather, within the set of physically acceptable approximants considered here and for the fiducial cosmology and accuracy criterion adopted in this work, it provides the largest usable redshift interval. Relative to the $z_{\max}\simeq0.70$ limit employed in Paper I, the resulting $z_{\max}\simeq1.61$ more than doubles the redshift range available to the cosmographic analysis.\\
\indent The redshift limit $z_{\max}=1.61$ is therefore adopted in the construction of the mock gravitational-wave bright-siren catalogues used throughout the remainder of this work. In the following subsection, we describe the source redshift distributions, the expected and optimistic numbers of events for LISA, ET, and DECIGO, and the procedure used to generate the simulated luminosity-distance measurements and their associated uncertainties.

\subsection{Gravitational Waves Source Distribution}
\label{sec:GW_distribution}

\begin{figure}[h]
    \centering
    \includegraphics[width=0.45\textwidth]{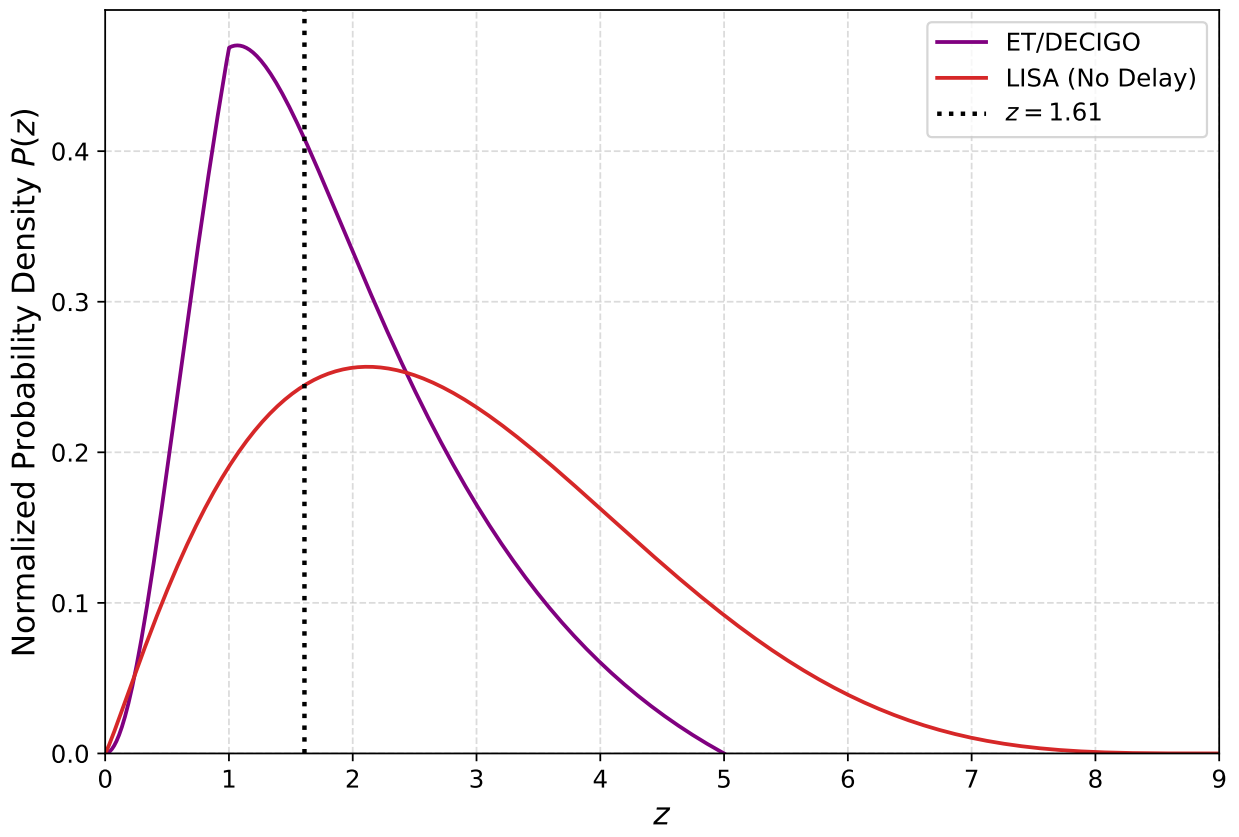}
    \caption{Normalized redshift probability density functions adopted for the gravitational-wave bright-siren mock catalogues. The ET and DECIGO samples follow the stellar-origin compact-binary distribution of Eq.~\eqref{P(z)_stellar_binaries}, while the LISA sample follows the No Delay massive-black-hole binary distribution of Eq.~\eqref{P(z)_LISA}. The vertical dotted line marks $z_{\max}=1.61$, corresponding to the adopted validity limit of the $P_{3,2}$ Padé parametrization.}
    \label{fig:redshift_distributions}
\end{figure}

\indent The redshift distribution of coalescing compact binaries provides the statistical prescription used to generate the mock gravitational-wave bright-siren catalogues. For the stellar-origin binary systems considered for the Einstein Telescope and DECIGO, the redshift probability density is written as~\cite{Zhao:2010sz}
\begin{equation}
\label{P(z)_stellar_binaries}
P(z)\propto\frac{4\pi d_C^2(z)R(z)}{H(z)(1+z)},
\end{equation}
where $d_C(z)=c\int_0^z H^{-1}(z')\,dz'$ is the comoving distance and $R(z)$ describes the evolution of the merger rate following the cosmic star-formation history~\cite{Schneider:2000sg}. The factor $4\pi d_C^2(z)$ accounts for the comoving-volume contribution, while the factor $(1+z)^{-1}$ incorporates the cosmological time dilation between the source and detector frames.\\
\indent The source population relevant for LISA is qualitatively different from that of ET and DECIGO. Rather than stellar-mass binaries, we consider massive black-hole binary (MBHB) mergers with identifiable electromagnetic counterparts, for which the GW luminosity distance can be combined with an independent redshift measurement of the host galaxy. These events therefore constitute bright standard sirens and represent one of the main classes of cosmological sources expected to be accessible to LISA in the millihertz band. Previous LISA forecasts describe this population through different astrophysical formation scenarios, including the Pop III, Q3d and Q3nod models~\cite{Cai:2017aea,Tamanini:2016}. In the present analysis, we adopt the Q3nod, or No Delay, corresponding to a heavy-seed population in which no delay is introduced between the merger of the host galaxies and the subsequent coalescence of their central massive black holes. Among the scenarios considered, Q3nod yields the largest expected event sample and is therefore adopted here as the most optimistic case for assessing the cosmological potential of LISA bright sirens.\\
\indent For numerical convenience, we describe the redshift distribution of the LISA bright-siren population in the No Delay scenario using the analytical fit introduced in Ref.~\cite{Ferreira:2022}, which reproduces the corresponding simulated MBHB distribution through
\begin{equation}
\label{P(z)_LISA}
P_{\rm LISA}(z)=\gamma\left(\frac{z}{9}\right)^{\alpha-1}\left(1-\frac{z}{9}\right)^{\beta-1},
\end{equation}
with
\begin{equation}
\alpha=2.14,\qquad\beta=4.70,\qquad\gamma=3.61.
\end{equation}
The fit is defined over the range relevant to the LISA MBHB sample, which extends to significantly higher redshifts than the stellar-origin sources considered for ET and DECIGO.\\
\indent We simulate bright-siren catalogues for three future gravitational-wave observatories: ET, DECIGO, and LISA. For ET and DECIGO, binary neutron-star systems are assumed to represent the relevant bright-siren population, provided that an electromagnetic counterpart allows the source redshift to be independently determined~\cite{Patricelli:2022hhr,Liu:2022mcd}. For ET, approximately $10^3$ bright sirens are expected over five years of operation, with sources detectable up to $z\simeq3$~\cite{ETDesignReport2020}. DECIGO is expected to detect of order $10^4$ bright sirens, with events extending to $z\simeq5$~\cite{Liu:2024gne}. For LISA, forecasts based on the No Delay MBHB population predict approximately $40$ potentially useful bright-siren events, consistent with previous estimates for the Q3nod scenario~\cite{Cai:2017aea}.\\
\indent The normalized redshift distributions adopted in this work for the corresponding mock catalogues are shown in Fig.~\ref{fig:redshift_distributions}. The stellar-origin population used for ET and DECIGO is more strongly concentrated at intermediate redshifts, whereas the LISA No Delay MBHB distribution is broader and extends to substantially higher redshifts.

\indent To determine the fraction of these populations that can be consistently included in the cosmographic analysis, we retain only sources within the validity domain of the $P_{3,2}$ parametrization. Events are therefore drawn from the corresponding distributions and selected according to $0<z<1.61$. Repeating this procedure over multiple realizations yields average effective samples of approximately $N_z=470$ events for ET, $N_z=4700$ for DECIGO, and $N_z=10$ for LISA. The adopted total populations and the corresponding numbers retained within the cosmographic range are summarized in Table~\ref{tab:N_initial}.

\begin{table}[ht]
\centering
\caption{Expected total number of events ($N_e$) over the full source population considered for each detector and effective number of events ($N_z$) within the adopted cosmographic range $z<1.61$.}
\label{tab:N_initial}
\begin{tabular}{lcc}
\hline
\textbf{Interferometer} & $\boldsymbol{N_e}$ & $\boldsymbol{N_z}$ \\
\hline
LISA                & 40    & 10   \\
Einstein Telescope  & 1000  & 470  \\
DECIGO              & 10000 & 4700 \\
\hline
\end{tabular}
\end{table}

\indent Compared with our previous analysis obtained in Paper I based on the third-order Taylor expansion, the extended redshift validity provided by the $P_{3,2}$ Padé parametrization substantially increases the number of bright sirens that can be consistently included in the cosmographic analysis. In our previous work, the restriction to $z\lesssim0.7$ resulted in effective samples of approximately $N_z=100$ for ET and $N_z=800$ for DECIGO, whereas in the present paper, the limit $z_{\max}=1.61$ increases these numbers to $N_z=470$ and $N_z=4700$, respectively. The broader redshift coverage therefore allows a significantly larger fraction of the expected source populations to contribute to the inference, while also enabling the inclusion of the LISA MBHB bright-siren population, with an effective sample of $N_z=10$. Having established the source populations and the number of events accessible within the adopted cosmographic domain, we now turn to the modelling of their gravitational-wave detection and the construction of the corresponding mock luminosity-distance measurements.

\subsection{Gravitational Waves Detection}

\indent Having specified the source populations and their redshift distributions, we now describe the procedure used to model the gravitational-wave signals and determine the corresponding luminosity-distance uncertainties. In the transverse-traceless (TT) gauge, the strain measured by an interferometric detector can be written as
\begin{equation}
h(t)=F_{+}(\theta,\phi,\psi)\,h_{+}(t)+F_{\times}(\theta,\phi,\psi)\,h_{\times}(t),
\end{equation}
where $F_{+}$ and $F_{\times}$ are the antenna pattern functions associated with the plus and cross polarizations, respectively~\cite{Sathyaprakash:2009xs}. The angular variables $(\theta,\phi)$ determine the source position in the detector frame, whereas $\psi$ denotes the polarization angle. The detector response depends on the interferometer geometry and therefore differs among ET, DECIGO, and LISA~\cite{Cai:2017aea,Califano:2022cmo,Yagi:2011wg}.\\
\indent In the frequency domain, the inspiral waveform is modeled using the stationary-phase approximation,
\begin{equation}
\label{phase_eq}
\mathcal{H}(f)=\mathcal{A}f^{-7/6}e^{i\Psi(f)},
\end{equation}
where the amplitude is
\begin{equation}
\label{amplitude}
\mathcal{A}=\frac{1}{d_L}\sqrt{F_+^2(1+\cos^2\iota)^2+4F_\times^2\cos^2\iota}\sqrt{\frac{5\pi}{96}}\,\pi^{-7/6}\mathcal{M}_c^{5/6}.
\end{equation}
Here, $d_L$ is the luminosity distance, $\iota$ is the inclination angle between the binary orbital angular momentum and the line of sight, and
\begin{equation}
\mathcal{M}_c=(1+z)M\eta^{3/5}
\end{equation}
is the observed chirp mass, where $M=m_1+m_2$ and $\eta=m_1m_2/M^2$.\\
\indent The simulated mass distributions depend on the source population. Following Ref.~\cite{Cai:2017aea}, for the stellar-origin binary neutron stars considered for ET and DECIGO, both component masses are drawn uniformly from the interval $[1,2]\,M_\odot$. For the LISA MBHB population, the component black-hole masses are sampled in the interval $[10^4,10^7]\,M_\odot$. The waveform phase $\Psi(f)$ is computed within the post-Newtonian approximation up to 3.5PN order~\cite{Nishizawa_2011}, while spin contributions are neglected.\\
\begin{figure*}[!t]
    \centering

    \begin{subfigure}{0.48\textwidth}
        \centering
        \includegraphics[width=\textwidth]{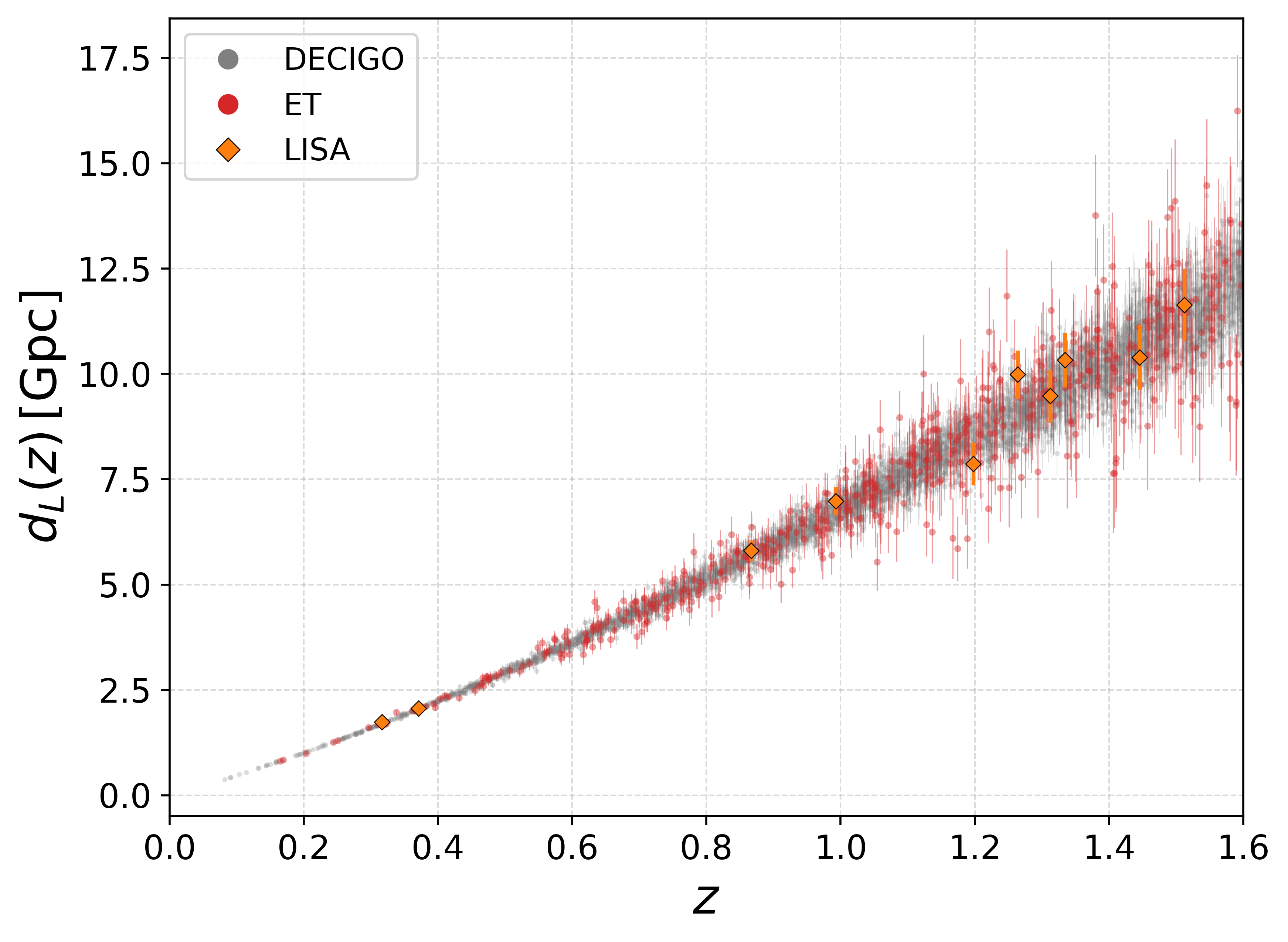}
        \label{fig:dl_sim}
    \end{subfigure}
    \hfill
    \begin{subfigure}{0.48\textwidth}
        \centering
        \includegraphics[width=\textwidth]{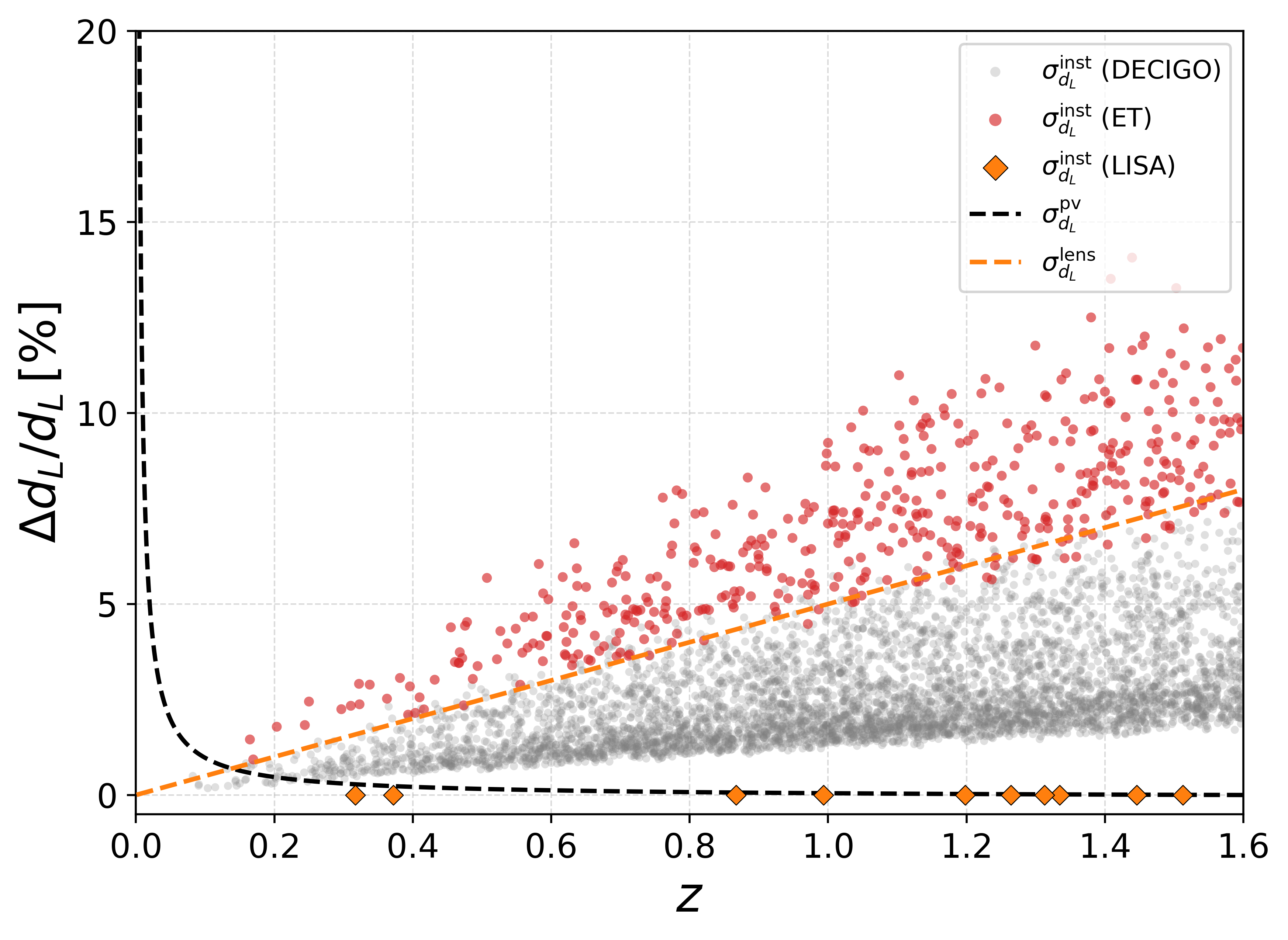}
        \label{fig:sigma_frac}
    \end{subfigure}

    \caption{
Simulated luminosity-distance catalogues for DECIGO, ET, and LISA, with the number of events corresponding to the expected detections adopted for each interferometer and restricted to the redshift interval considered in the analysis (left panel). Fractional uncertainties on the luminosity distance, where the points show the instrumental contribution, $\sigma_{d_L}^{\rm inst}/d_L$, obtained from the simulated events through their individual signal-to-noise ratios, while the dashed curves represent the contributions from peculiar velocities and weak gravitational lensing. All contributions are expressed as a percentage of the luminosity distance (right panel).
}

    \label{fig:simulations_erros}
\end{figure*}
\indent The treatment of the detector response requires an additional consideration for LISA. Since MBHB inspirals can remain in the LISA sensitivity band for several months, the orientation of the source with respect to the detector changes appreciably during the observation as the LISA constellation follows its orbit around the Sun. We therefore account for the time dependence of the LISA antenna pattern functions following the prescription of Ref.~\cite{Cai:2017aea}, in which the source angles entering $F_{+}$ and $F_{\times}$ evolve with the orbital motion of the constellation. Following the same prescription, no analogous time-dependent modulation is introduced for the ET and DECIGO responses adopted in our simulations.\\
\indent Once the waveform is specified, the detectability of each simulated event is quantified through the network signal-to-noise ratio (SNR). A source is retained only if
\begin{equation}
\rho>8,
\end{equation}
with the combined network SNR defined as
\begin{equation}
\label{SNR}
\rho=\sqrt{\sum_{i=1}^{N}\left(\rho^{(i)}\right)^2},
\end{equation}
where $\rho^{(i)}=\sqrt{\langle\mathcal{H}^{(i)},\mathcal{H}^{(i)}\rangle}$ is the SNR associated with the $i$th independent interferometric channel. The noise-weighted inner product is
\begin{equation}
\langle a,b\rangle\equiv4\int_{f_{\min}}^{f_{\max}}\frac{\tilde{a}(f)\tilde{b}^{*}(f)+\tilde{a}^{*}(f)\tilde{b}(f)}{2}\frac{df}{S_h(f)},
\end{equation}
where $S_h(f)$ denotes the one-sided noise power spectral density (PSD), which characterizes the frequency-dependent noise level of the detector.\\
\indent For ET, we adopt the ET-D sensitivity configuration~\cite{Hild:2010id}, while the DECIGO PSD is implemented following Ref.~\cite{Liu:2024gne}. For LISA, the detector response and sensitivity prescription are implemented following Ref.~\cite{Cai:2017aea}, including the orbital modulation of the antenna pattern and the corresponding millihertz sensitivity band. The lower-frequency cutoff is taken to be $f_{\min}=10^{-4}$ Hz, while the upper cutoff is determined by the detector response at high frequency. Events that do not satisfy the threshold $\rho>8$ are rejected and resampled until the required catalogue size is reached.\\
\indent After selecting the detectable events, we estimate the uncertainty on the luminosity distance. The instrumental contribution is obtained from the Fisher-matrix approximation,
\begin{equation}
\sigma^{\rm inst}_{d_L}\simeq\sqrt{\left\langle\frac{\partial\mathcal{H}}{\partial d_L},\frac{\partial\mathcal{H}}{\partial d_L}\right\rangle^{-1}}.
\end{equation}
\indent Since $\mathcal{H}\propto d_L^{-1}$, this reduces approximately to $\sigma^{\rm inst}_{d_L}\simeq d_L/\rho$. To conservatively account for the correlation between luminosity distance and inclination, we adopt
\begin{equation}
\label{inst_error}
\sigma^{\rm inst}_{d_L}\simeq\frac{2d_L}{\rho},
\end{equation}
following Ref.~\cite{Cai:2017aea}.\\
\indent The weak-lensing contribution is modeled as~\cite{Zhao:2010sz}
\begin{equation}
\sigma^{\rm lens}_{d_L}=0.05\,z\,d_L,
\end{equation}
while the uncertainty associated with peculiar velocities is given by
\begin{equation}
\sigma^{\rm pv}_{d_L}(z)=d_L(z)\left|1-\frac{(1+z)^2}{H(z)d_L(z)}\right|\sigma_{v,\rm gal},
\end{equation}
with $\sigma_{v,\rm gal}=300\,{\rm km\,s^{-1}}$~\cite{Gordon:2007zw}.\\
\indent The total uncertainty on the luminosity distance is obtained by adding the individual contributions in quadrature,
\begin{equation}
\label{total_dl_error}
\sigma_{d_L}(z)=\sqrt{\left[\sigma^{\rm inst}_{d_L}(z)\right]^2+\left[\sigma^{\rm lens}_{d_L}(z)\right]^2+\left[\sigma^{\rm pv}_{d_L}(z)\right]^2}.
\end{equation}
\indent Since we restrict the analysis to bright sirens with identified electromagnetic counterparts, the redshift uncertainty is neglected. Finally, mock luminosity-distance measurements are generated by drawing each event from a Gaussian distribution centered on the fiducial luminosity distance, with standard deviation $\sigma_{d_L}(z)$.\\
\indent The simulated luminosity-distance catalogues are shown in the left panel of Fig.~\ref{fig:simulations_erros}. The larger numbers of DECIGO and ET events provide a dense sampling of the luminosity-distance--redshift relation over the full redshift interval considered, while the smaller LISA bright-siren sample appears as a sparse set of individual events. The dispersion of the simulated measurements increases toward higher redshifts, reflecting the growth of the luminosity-distance uncertainties. Despite the smaller number of LISA events, their individual measurements remain well constrained because of their very large signal-to-noise ratios.\\
\indent The right panel of Fig.~\ref{fig:simulations_erros} shows the different contributions to the luminosity-distance uncertainty. As expected, the peculiar-velocity contribution dominates the error budget at low redshifts, where even small deviations from the ideal Hubble flow can represent a significant fraction of the inferred luminosity distance~\cite{Blake:2025etn}. At higher redshifts, this contribution rapidly decreases, while weak gravitational lensing becomes progressively more relevant due to the cumulative distortion of the GW propagation path by the intervening large-scale structure of the Universe~\cite{Canevarolo:2023dkh}. The instrumental contribution depends on the detector sensitivity and on the SNR of each event. ET exhibits larger instrumental uncertainties than DECIGO over most of the considered redshift range, whereas the instrumental contribution for the LISA events is practically negligible. This behavior is a direct consequence of the very high SNR values of the selected LISA MBHB events, for which $\sigma^{\rm inst}_{d_L}\propto\rho^{-1}$ becomes much smaller than the other sources of uncertainty. Consequently, the luminosity-distance uncertainty of the LISA bright sirens is mainly determined by weak gravitational lensing, while peculiar velocities remain relevant only at sufficiently low redshifts.

\subsection{Parameter Estimation} \label{sec: param_est}

In this work, we adopt a Bayesian framework~\cite{Trotta:2017wnx} to estimate the parameters $\{\theta_i\} = \{H_0, q_0, j_0, s_0, l_0\}$ given the model $\mathcal{M}$, represented by the cosmographic expansions in Eq.~\eqref{p32}. In this context, the joint posterior probability distribution of the parameters $\theta_i$ is given by
\begin{equation}
P(\{\theta_i\}|d, \mathcal{M}) = \frac{P(d|\{\theta_i\}, \mathcal{M})\, P(\{\theta_i\}|\mathcal{M})}{P(d)},
\end{equation}
where the likelihood $P(d|\{\theta_i\}, \mathcal{M})$ is defined as
\begin{equation}
P(d|\{\theta_i\}, \mathcal{M}) = \exp \left[ -\frac{1}{2} \sum_{i=1}^{N} 
\frac{|d_i - f_i(\{\theta_i\}, z_i)|^2}{\sigma_i^2} \right],
\end{equation}
with $N$ being the number of detections, $d_i$ the observed data with uncertainty $\sigma_i$, and $f_i(\{\theta_i\}, z_i)$ the model prediction.\\
\indent The priors on $H_0$, $q_0$, and $j_0$ are assumed to be uniform within the ranges $[54,81]\,\mathrm{km\,\cdot\,s^{-1}\,\cdot\,Mpc^{-1}}$, $[-1,0]$, and $[-1,3]$, respectively. For the higher-order cosmographic parameters, we adopt uniform priors $s_0 \in [-10,10]$ and $l_0 \in [-30,30]$. These intervals are chosen to encompass current observational constraints while remaining sufficiently broad to avoid informative priors~\cite{Morais:2026qdw, deSouza:2021xtg, Capozziello:2020ctn,  Hu:2022udt}. We note that the uncertainty in the redshift is neglected, consistently with the assumption of an electromagnetic counterpart to the gravitational wave signal, which allows for the unambiguous identification of the host galaxy.

\begin{figure*}[t!]
\centering

\begin{subfigure}{\textwidth}
\centering
\includegraphics[width=0.65\textwidth]{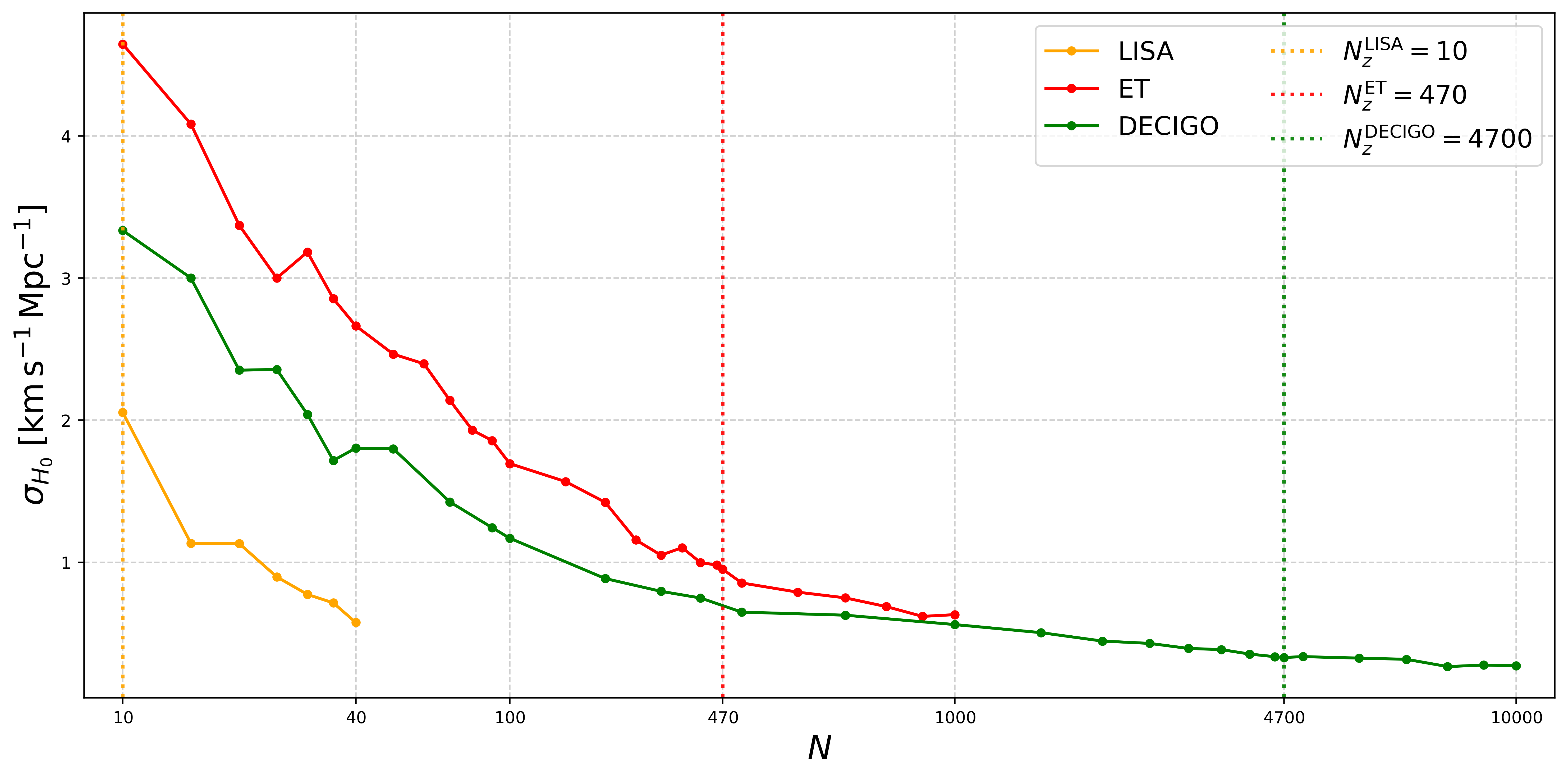}
\label{fig:sigma_H0_N}
\end{subfigure}


\begin{subfigure}{\textwidth}
\centering
\includegraphics[width=0.65\textwidth]{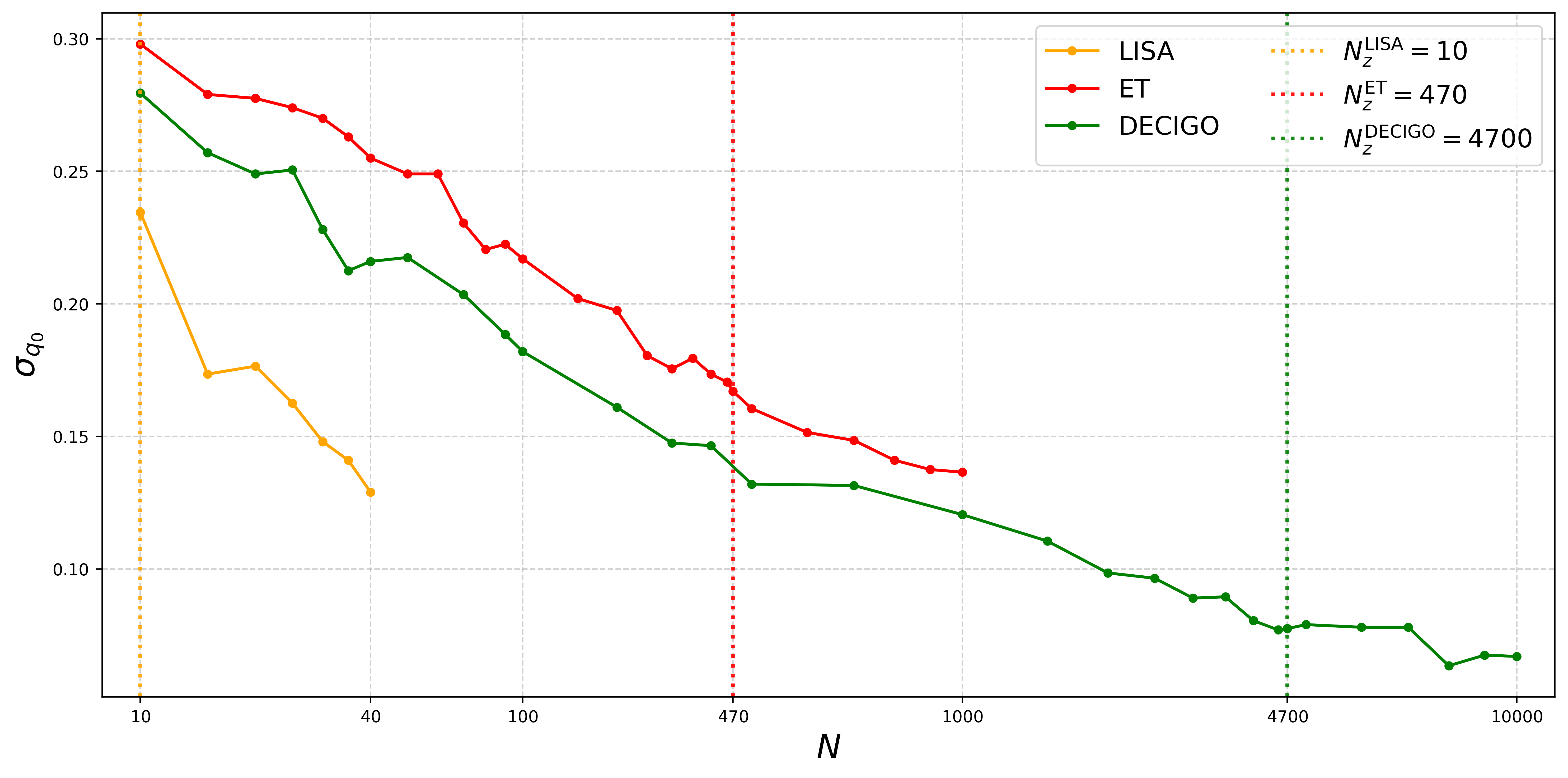}
\label{fig:sigma_q0_N}
\end{subfigure}


\begin{subfigure}{\textwidth}
\centering
\includegraphics[width=0.65\textwidth]{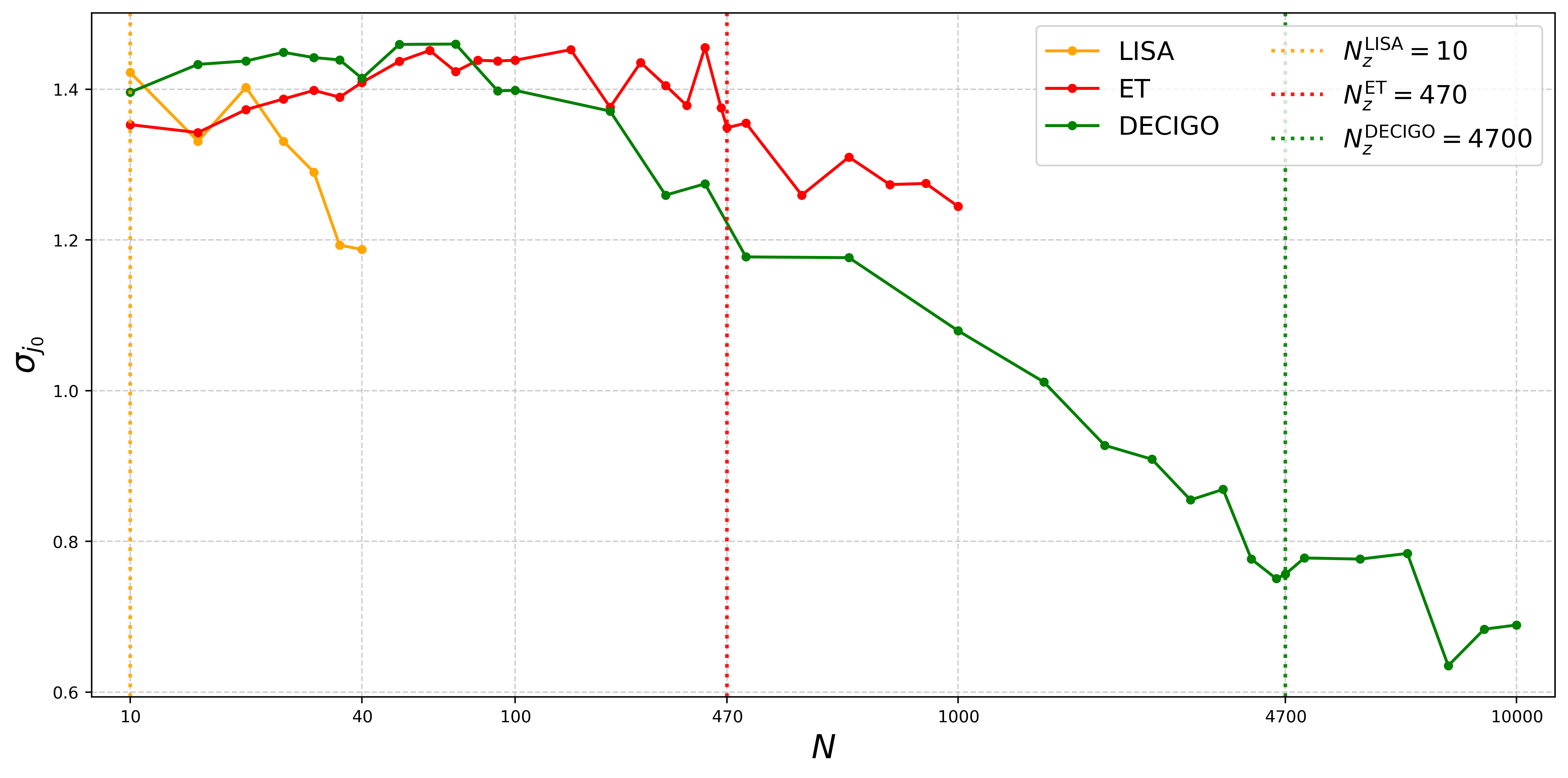}
\label{fig:sigma_j0_N}
\end{subfigure}

\caption{The top, middle, and bottom panels show, respectively, the correspond uncertainties of the cosmographic parameters $H_0$, $q_0$, and $j_0$ as functions of the number of detections, $N$. The curves show the results obtained for LISA, the Einstein Telescope, and DECIGO. The vertical dotted lines indicate the reference catalogue sizes adopted for each interferometer: $N=10$, $470$, and $4700$, respectively.}

\label{fig:sigma_N_Hqj}
\end{figure*}
\begin{figure*}[!t]
    \centering

    \begin{subfigure}{0.49\textwidth}
        \centering
        \includegraphics[width=\textwidth]{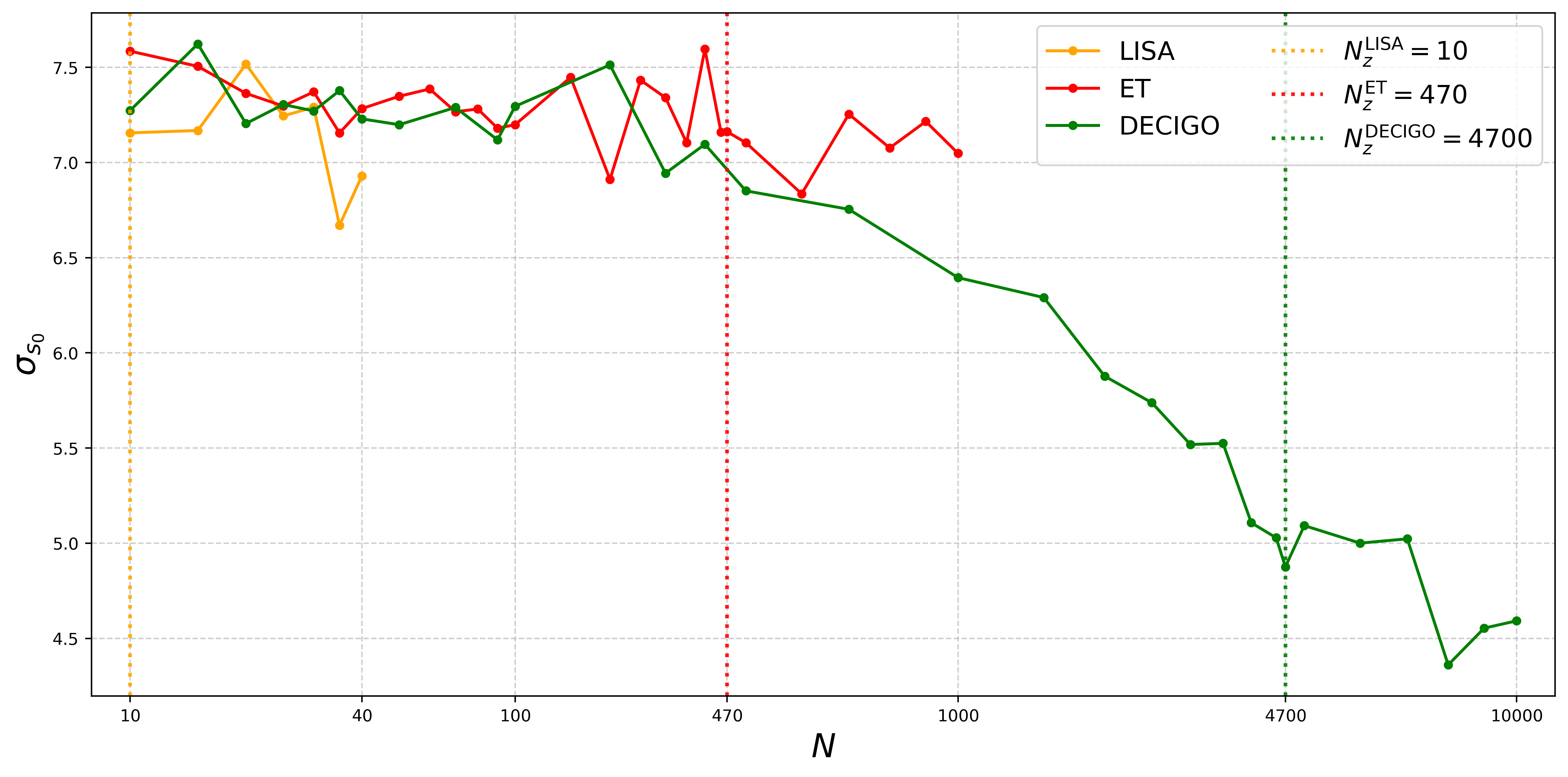}
        \label{fig:sigma_s0_N}
    \end{subfigure}
    \hfill
    \begin{subfigure}{0.49\textwidth}
        \centering
        \includegraphics[width=\textwidth]{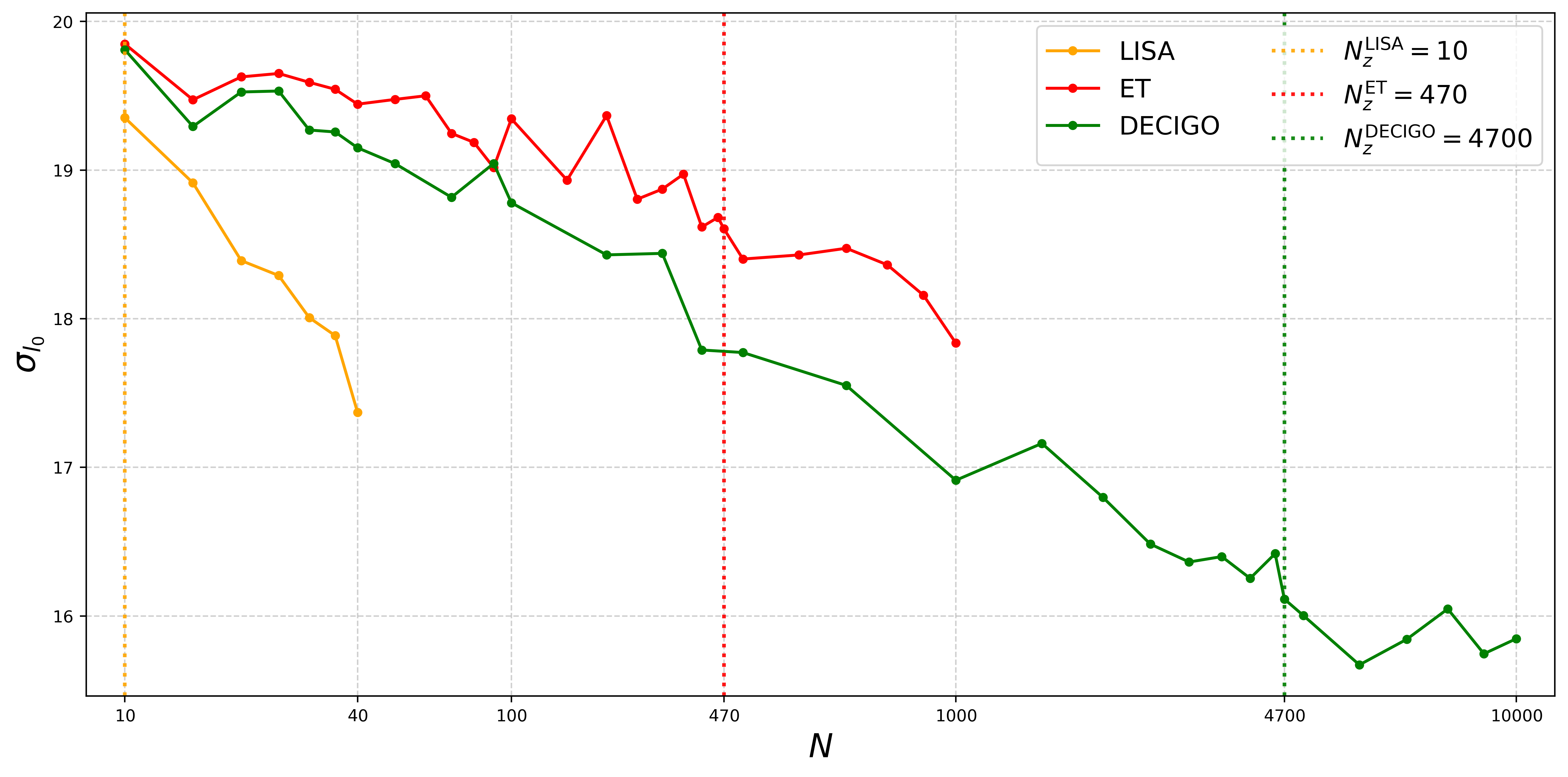}
        \label{fig:sigma_l0_N}
    \end{subfigure}

    \caption{Uncertainties, $\sigma$, of the cosmographic parameters $s_0$ and $l_0$ as functions of the number of detections, $N$. The left and right panels correspond to $s_0$ and $l_0$, respectively. The curves show the results obtained for LISA, the Einstein Telescope, and DECIGO. The vertical dotted lines indicate the reference catalogue sizes adopted for each interferometer: $N=10$, $470$, and $4700$, respectively.}

    \label{fig:sigma_N_sl}
\end{figure*}

\section{Results} \label{sec:results}

\indent In this section, we apply the methodology described above to investigate the cosmographic constraints achievable with the $P_{3,2}$ Padé parametrization using simulated bright-siren catalogues from LISA, ET, and DECIGO. By extending the analysis to $z\leq1.6$, the Padé approach allows us to exploit a substantially broader redshift range than the low-redshift Taylor expansion considered in our previous analysis, thereby incorporating a larger fraction of the events expected from future gravitational-wave observatories. As in our previous work, we consider two complementary approaches based on the number of detections ($N$) and the signal-to-noise ratio (SNR). The analysis as a function of $N$ constitutes our main investigation and quantifies the improvement in the cosmographic constraints as progressively larger catalogues become available, whereas the SNR-based analysis is used to examine how the precision of the inferred parameters depends on the properties and redshift distribution of the detected signals.\\
\indent For each bin in both analysis, we performed 50 independent realizations and estimated the average of the mean values and the average of the standard deviations across all realizations. The posterior distributions of $H_0$, $q_0$, $j_0$, $s_0$ and $l_0$ were obtained through a Markov Chain Monte Carlo (MCMC) approach using \texttt{emcee}~\cite{Foreman_Mackey_2013}. A comprehensive summary of the adopted values of $N$, the SNR intervals, and the MCMC-derived cosmographic results is presented in~\ref{appendix:N} and~\ref{appendix:SNR}. In addition, Figures~\ref{fig:sigma_N_Hqj},~\ref{fig:sigma_N_sl}, ~\ref{fig:sigma_SNR_Hqj} and~\ref{fig:sigma_SNR_sl} illustrate the behavior of the parameter uncertainties as functions of $N$ and SNR, respectively, for all interferometers analyzed.

\subsection{$N$ Analysis}

\indent In the first part of the analysis, we investigate how the cosmographic constraints evolve as the number of detected events increases. The three interferometers are explored over the common interval $10\leq N\leq40$, allowing their constraints to be compared at fixed catalogue sizes. ET and DECIGO are additionally evaluated at common values of $N$ up to $N=1000$, while the DECIGO analysis is extended to $N=10000$. The reference catalogue sizes within the adopted redshift interval, $N_z=10$, $470$, and $4700$ for LISA, ET, and DECIGO, respectively, are indicated by the vertical dotted lines in Figs.~\ref{fig:sigma_N_Hqj} and~\ref{fig:sigma_N_sl}. These values represent the expected number of detections adopted for each experiment within $z\leq1.6$. Catalogue sizes below $N_z$ can therefore be interpreted as progressively earlier stages of event accumulation, whereas values above $N_z$ represent increasingly optimistic scenarios in which larger samples become available. Comparisons among the detectors are performed only at common values of $N$, while the remaining portions of the curves illustrate the evolution of the constraints for each experiment individually.\\
\indent The top and middle panels of Fig.~\ref{fig:sigma_N_Hqj} show that the uncertainties on $H_0$ and $q_0$ decrease as the catalogue size grows for all three interferometers, with the improvement being most pronounced for $H_0$. At the reference catalogue sizes, the uncertainties on the Hubble constant are approximately $2.05$, $0.95$, and $0.33~\mathrm{km\,s^{-1}\,Mpc^{-1}}$ for LISA, ET, and DECIGO, respectively, corresponding to percent-level or subpercent constraints. For $q_0$, the corresponding uncertainties are approximately $0.235$, $0.167$, and $0.078$. The overall decrease continues toward the largest catalogues considered, although the gain becomes progressively more gradual at large $N$.\\
\indent The behavior of the higher-order parameters is less pronounced. As shown in the bottom panel of Fig.~\ref{fig:sigma_N_Hqj}, the uncertainty on $j_0$ decreases more slowly and exhibits fluctuations as the catalogue size increases, particularly for ET and DECIGO. At the reference values $N_z$, the corresponding uncertainties are approximately $1.42$, $1.35$, and $0.76$ for LISA, ET, and DECIGO, respectively. The same general pattern extends to $s_0$ and $l_0$, shown in Fig.~\ref{fig:sigma_N_sl}: their constraints improve only modestly over much of the explored range and remain substantially broader than those obtained for the lower-order parameters. At the reference catalogue sizes, $\sigma_{s_0}$ is approximately $7.16$, $7.16$, and $4.88$, while $\sigma_{l_0}$ is approximately $19.35$, $18.60$, and $16.11$ for LISA, ET, and DECIGO, respectively. The improvement becomes more evident for the largest DECIGO samples, although both parameters remain weakly constrained compared with $H_0$ and $q_0$.\\
\indent Overall, increasing the number of detections leads to tighter cosmographic constraints, but the rate of improvement decreases as progressively higher-order parameters are considered. The complete numerical results for all catalogue sizes are reported in Tables~\ref{tab:LISA_N}, \ref{tab:ET_N}, and~\ref{tab:DECIGO_N}.

\subsection{SNR Analysis}

\begin{figure}[t]
\centering
\includegraphics[width=0.45\textwidth]{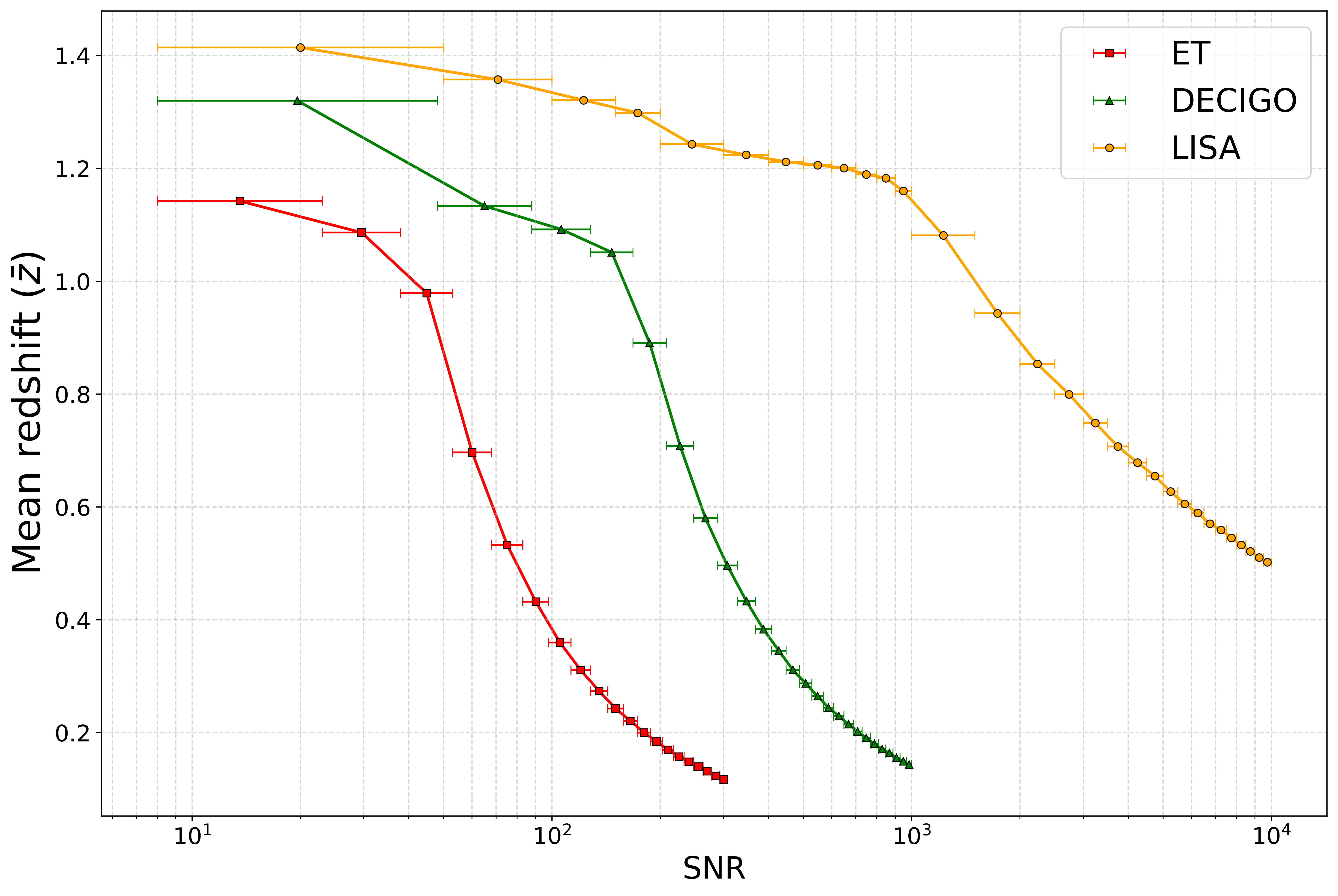}
\caption{Mean redshift of the mock catalogues as a function of SNR for ET (red circles), DECIGO (green circles), and LISA (orange circles). Each marker is positioned at the central SNR value of its corresponding interval, while the horizontal segments represent the complete detector-specific SNR bins. The results are averaged over 50 realizations with $N=100$ events per bin. The decrease in $\bar{z}$ toward higher SNR illustrates the correlation between these quantities.}
\label{fig:z_mean_SNR}
\end{figure}

\indent As a complementary part of our analysis, we investigate how the cosmographic constraints behave across different signal-to-noise ratio regimes. Unlike the previous analysis, in which the catalogue size was progressively increased, here we fix the number of events at $N=100$ for LISA, ET, and DECIGO in every SNR interval considered. For each detector-specific bin, source parameters are sampled until exactly 100 accepted events are obtained, and the procedure is repeated over 50 independent realizations. Keeping the catalogue size fixed prevents changes in the number of detections from driving the evolution of the uncertainties and allows us to examine how the inferred constraints vary as different portions of the SNR distribution are selected. These catalogues should therefore be interpreted as controlled statistical samples rather than as predictions for a common observing time or for the expected number of detections of each experiment. Since the three interferometers operate over substantially different SNR regimes, the intervals are defined separately for each detector.\\
\indent Although the number of events is fixed, the SNR selection also modifies the redshift distribution of the corresponding catalogues. Higher-SNR events tend, on average, to be associated with lower redshifts, so SNR and redshift cannot be treated as independent quantities in this analysis. To characterize this relation explicitly, we compute the mean redshift, $\bar{z}$, associated with each SNR interval, as shown in Fig.~\ref{fig:z_mean_SNR}. The systematic decrease of $\bar{z}$ toward higher SNR is observed for all three interferometers, although each detector spans a different SNR--redshift domain. Consequently, moving across the SNR intervals simultaneously changes both the typical signal quality and the redshift coverage of the mock catalogues. The following results must therefore be interpreted as the joint evolution of the cosmographic constraints with SNR and redshift at fixed $N$, rather than as the isolated effect of SNR alone. For the same reason, the curves should not be regarded as a direct ranking of the final observational precision of the three experiments.\\
\begin{figure*}[t!]
\centering
\begin{subfigure}{\textwidth}
\centering
\includegraphics[width=0.61\textwidth]{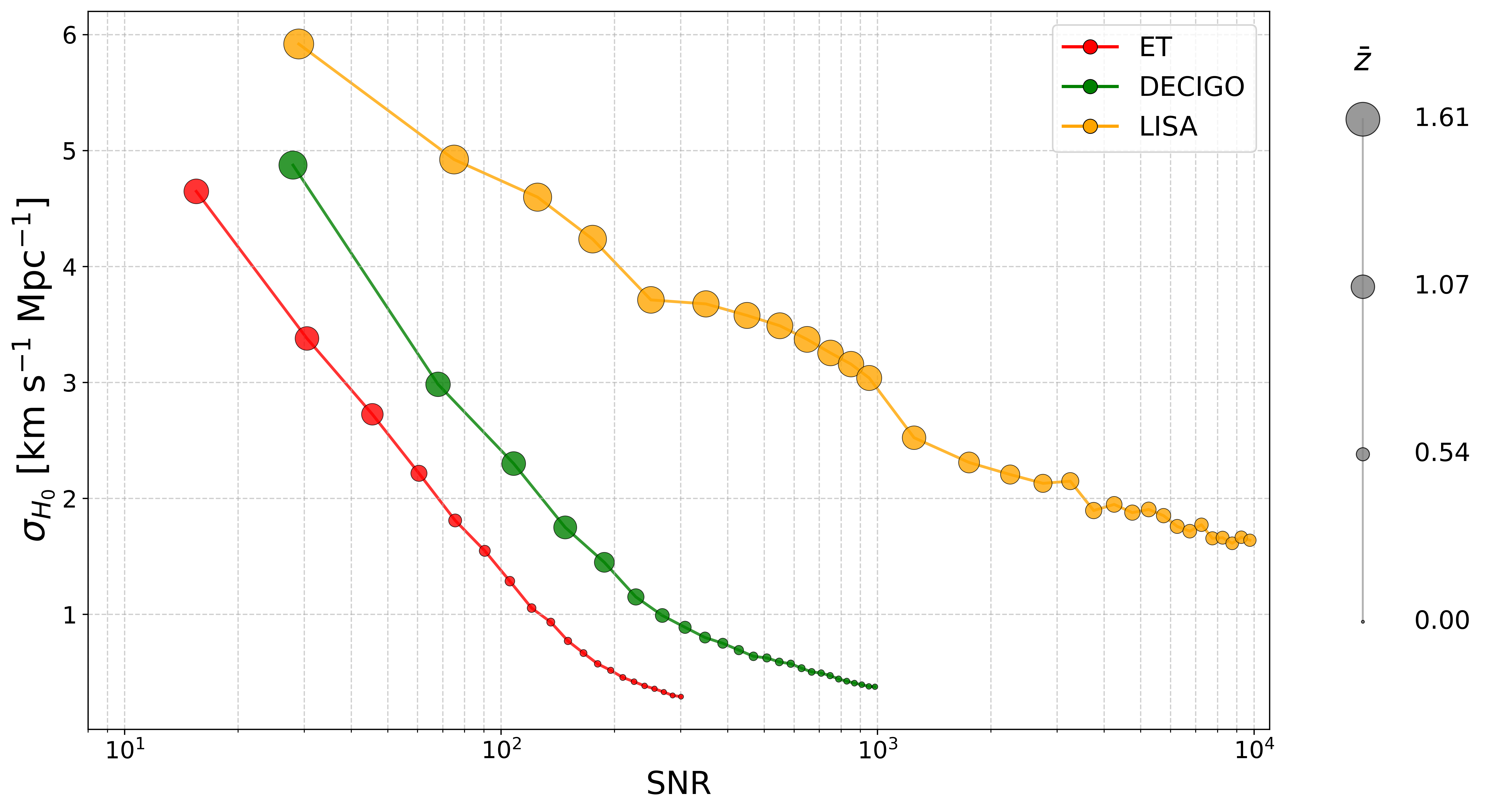}
\label{fig:sigma_H0_SNR}
\end{subfigure}
\begin{subfigure}{\textwidth}
\centering
\includegraphics[width=0.64\textwidth]{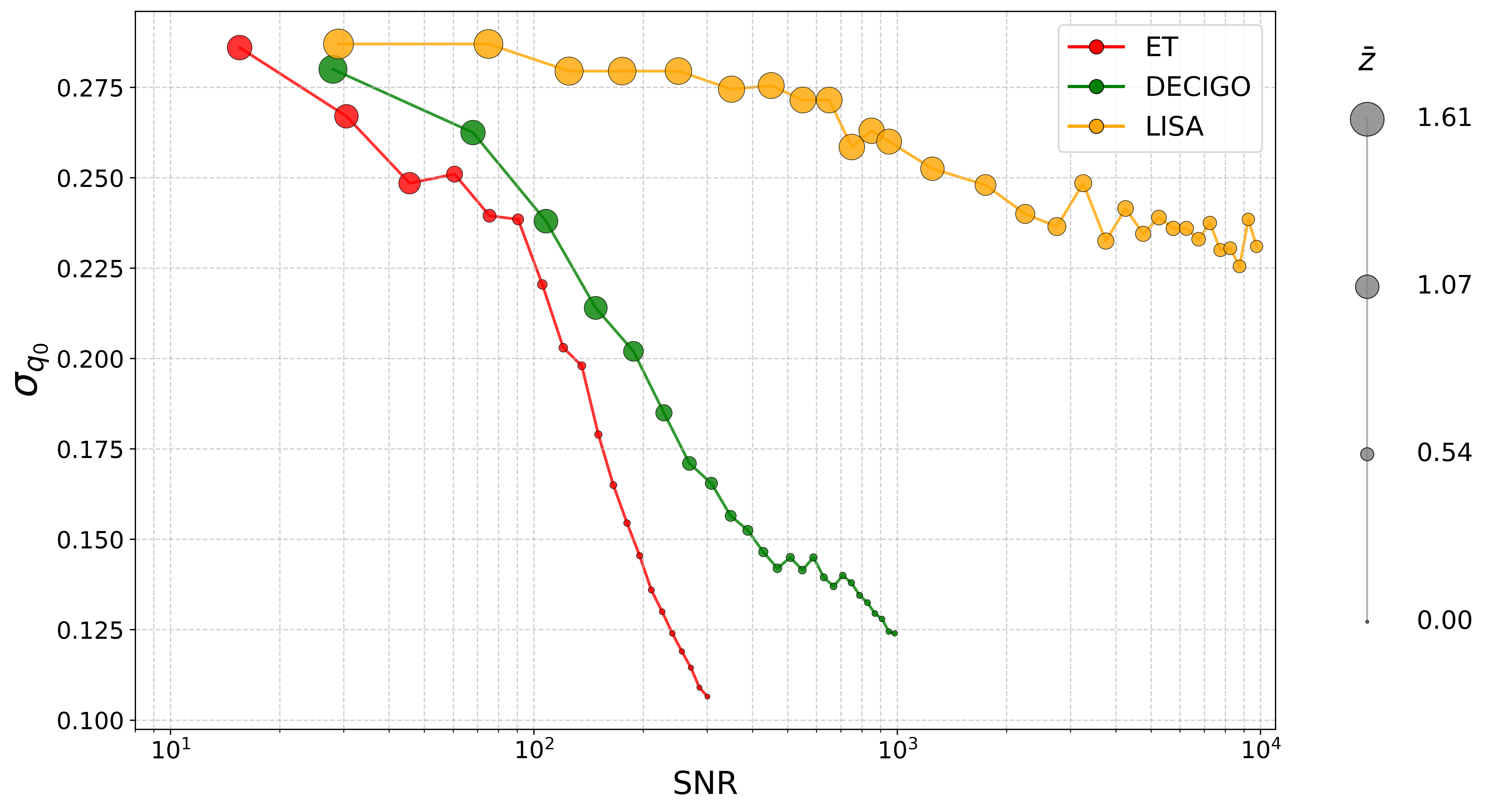}
\label{fig:sigma_q0_SNR}
\end{subfigure}
\begin{subfigure}{\textwidth}
\centering
\includegraphics[width=0.64\textwidth]{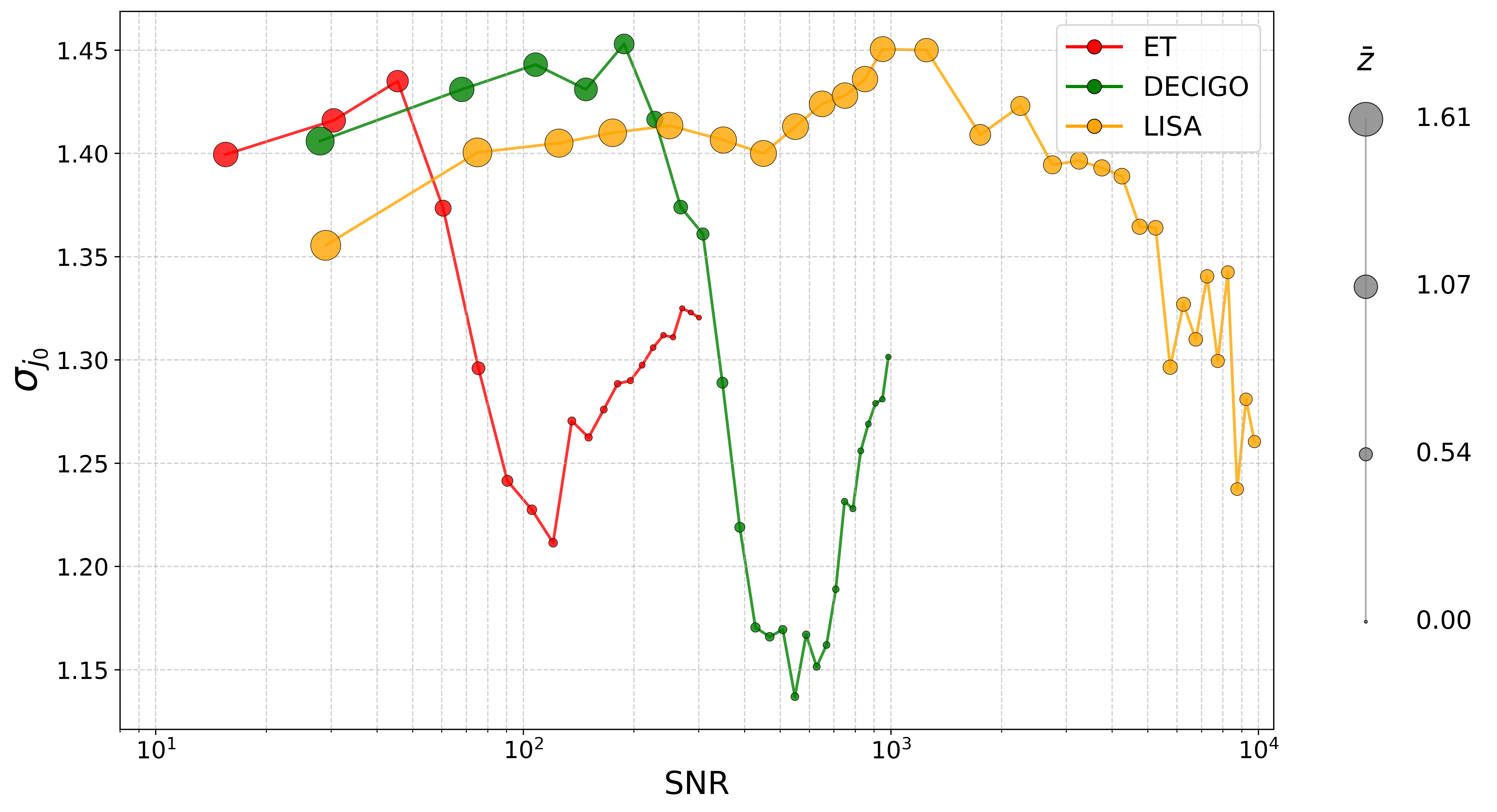}
\label{fig:sigma_j0_SNR}
\end{subfigure}
\caption{Uncertainties $\sigma$ of the cosmographic parameters $H_0$, $q_0$, and $j_0$ as functions of the signal-to-noise ratio. The top, middle, and bottom panels correspond to $H_0$, $q_0$, and $j_0$, respectively. Each point is positioned at the central SNR value of its corresponding interval, while its size represents the associated mean redshift, $\bar{z}$. Larger circles therefore correspond to higher-redshift catalogues.}
\label{fig:sigma_SNR_Hqj}
\end{figure*}
\indent With the SNR--redshift relation established, Fig.~\ref{fig:sigma_SNR_Hqj} shows that the uncertainties on $H_0$ and $q_0$ generally decrease toward the higher-SNR and lower-redshift regime. The trend is particularly clear for ET and DECIGO, while LISA exhibits a more gradual evolution over its substantially broader SNR range. For $H_0$, the highest-SNR bins reach uncertainties of approximately $0.3$--$0.4~\mathrm{km\,s^{-1}\,Mpc^{-1}}$ for ET and DECIGO, whereas LISA reaches approximately $1.6~\mathrm{km\,s^{-1}\,Mpc^{-1}}$.\\
\indent The behavior of $j_0$ is qualitatively different. Rather than improving continuously toward higher SNR, the ET and DECIGO curves reach their smallest uncertainties at intermediate SNR--redshift regimes and subsequently increase as the catalogues become increasingly concentrated at low redshift. For ET, the minimum occurs around $\mathrm{SNR}\in[113,128]$, with $\bar{z}\simeq0.31$, while for DECIGO it occurs around $\mathrm{SNR}\in[528,568]$, with $\bar{z}\simeq0.27$. LISA shows milder variations throughout its SNR range, without the same pronounced minimum. This behavior indicates that the gain associated with higher signal quality does not translate monotonically into tighter constraints on the higher-order expansion terms when the corresponding redshift leverage is simultaneously reduced.\\
\begin{figure*}[!t]
\centering
\begin{subfigure}{0.49\textwidth}
\centering
\includegraphics[width=\textwidth]{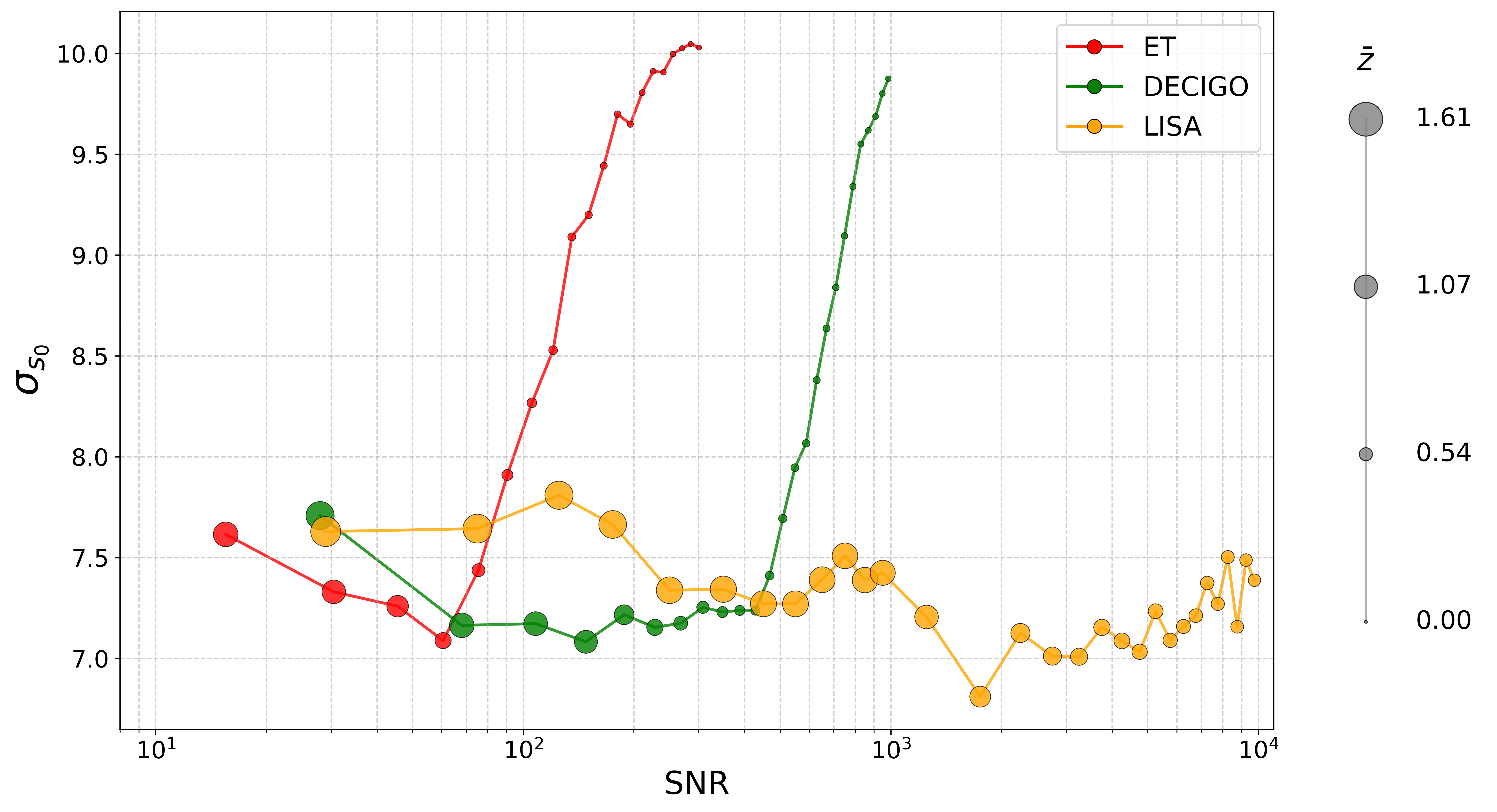}
\label{fig:sigma_s0_SNR}
\end{subfigure}
\hfill
\begin{subfigure}{0.49\textwidth}
\centering
\includegraphics[width=\textwidth]{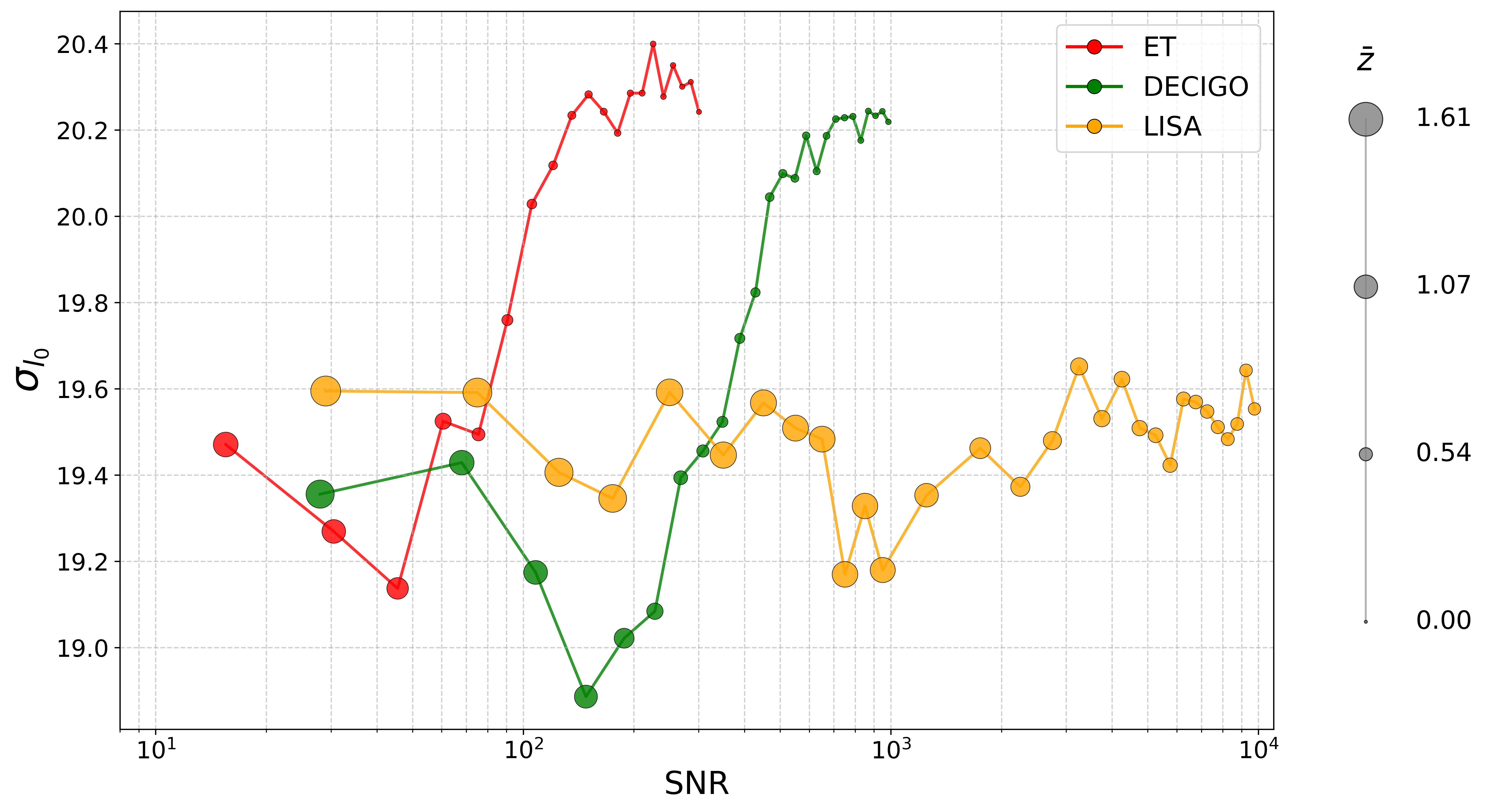}
\label{fig:sigma_l0_SNR}
\end{subfigure}
\caption{Uncertainties $\sigma$ of the cosmographic parameters $s_0$ and $l_0$ as functions of the signal-to-noise ratio. The left and right panels correspond to $s_0$ and $l_0$, respectively. Each point is positioned at the central SNR value of its corresponding interval, while its size represents the associated mean redshift, $\bar{z}$.}
\label{fig:sigma_SNR_sl}
\end{figure*}
\indent A similar dependence on the SNR--redshift regime is observed for the higher-order parameters in Fig.~\ref{fig:sigma_SNR_sl}. For ET and DECIGO, the uncertainties on $s_0$ and $l_0$ initially decrease or remain approximately stable before increasing toward the highest-SNR and lowest-redshift intervals. The LISA curves remain comparatively flatter, with only moderate variations over most of the explored range. Taken together with the behavior of $j_0$, these results emphasize the increasing importance of redshift coverage for constraining progressively higher orders of the cosmographic expansion. The complete numerical results for all SNR intervals and their associated mean redshifts are reported in Tables~\ref{tab:LISA_SNR}, \ref{tab:ET_SNR}, and~\ref{tab:DECIGO_SNR}.

\section{Conclusions} \label{sec: conclusion}

\indent This work extends gravitational-wave cosmography beyond the redshift range explored with the truncated Taylor expansion by adopting Padé representations of the luminosity-distance relation. After examining different rational approximants and requiring both a physically admissible behavior and agreement with the fiducial flat $\Lambda$CDM luminosity distance at the $1\%$ level, we selected the $P_{3,2}$ parametrization for the statistical analysis. Its validity up to $z\simeq1.61$ allows the simulated catalogues to include a substantially broader population of bright standard sirens from LISA, ET, and DECIGO. Within this framework, the expansion history is described through the five parameters $H_0$, $q_0$, $j_0$, $s_0$, and $l_0$, and the forecasts are evaluated primarily as a function of the catalogue size, with an additional fixed-$N$ analysis used to characterize their behavior across different SNR--redshift regimes.\\
\indent The catalogue-size analysis shows that the lowest-order parameters can be determined with substantially smaller uncertainties than the higher derivatives of the expansion. For the reference samples $N_z=10$, $470$, and $4700$ adopted for LISA, ET, and DECIGO, respectively, the relative uncertainty on $H_0$ is approximately $3.0\%$, $1.4\%$, and $0.5\%$, while the corresponding values for $q_0$ are approximately $44.0\%$, $31.3\%$, and $14.5\%$. Controlled comparisons at equal catalogue sizes also illustrate the role of the information carried by the individual events: at $N=40$, the relative uncertainties on $H_0$ are approximately $0.9\%$, $3.9\%$, and $2.7\%$ for LISA, ET, and DECIGO, respectively, while ET and DECIGO reach approximately $0.9\%$ and $0.8\%$ at $N=1000$. In the largest DECIGO catalogue considered, $N=10000$, the relative uncertainty on $H_0$ reaches approximately $0.4\%$, whereas that on $q_0$ is approximately $12.6\%$. These forecasts preserve the potential of standard sirens to provide an independent determination of the present expansion rate and therefore to contribute to the investigation of the Hubble tension~\cite{DiValentino:2021izs, Hu:2023jqc, Capozziello:2024stm, Kraiselburd:2025gti}. Since all mocks are generated around $H_0=67.66~\mathrm{km\,s^{-1}\,Mpc^{-1}}$, however, the analysis quantifies the achievable uncertainty without selecting either of the values currently in tension.\\
\indent Moving to higher cosmographic orders, the constraints become progressively broader. At the reference catalogue sizes, the relative uncertainty on $j_0$ is approximately $142\%$ for LISA, $135\%$ for ET, and $76\%$ for DECIGO. For the common value $N=1000$, ET and DECIGO yield uncertainties of approximately $124\%$ and $108\%$, respectively, while the DECIGO extension to $N=10000$ reaches approximately $69\%$. The still higher-order parameters remain substantially less constrained: at the reference samples, $\sigma_{s_0}\simeq7.16$, $7.16$, and $4.88$, and $\sigma_{l_0}\simeq19.35$, $18.60$, and $16.11$ for LISA, ET, and DECIGO, respectively. Even in the $N=10000$ DECIGO catalogue, the corresponding uncertainties remain at approximately $\sigma_{s_0}\simeq4.59$ and $\sigma_{l_0}\simeq15.85$. Thus, within the five-parameter $P_{3,2}$ description, the increasing cosmographic order is accompanied by progressively wider posterior intervals. The recovered values remain compatible with the fiducial cosmology used in the simulations, including the $\Lambda$CDM prediction $j_0=1$.\\
\indent The complementary SNR analysis was designed to isolate the evolution of the constraints across different signal-quality regimes while keeping the catalogue size fixed at $N=100$. Since higher-SNR bins systematically correspond to lower mean redshifts, the curves represent a combined evolution in SNR and redshift rather than the effect of SNR alone. The uncertainties on $H_0$ decrease steadily toward the high-SNR regime for all three interferometers, and $q_0$ follows the same overall tendency. By contrast, the higher-order quantities exhibit more complex behavior. For ET and DECIGO, the $j_0$, $s_0$, and $l_0$ curves develop minima at intermediate SNRs and subsequently increase toward the highest-SNR, lowest-redshift samples, while LISA displays comparatively milder variations over much of its much broader SNR range. Because every bin contains the same imposed number of events, these results are not forecasts of detector performance, but rather characterize how the cosmographic information changes as signal quality and redshift leverage vary simultaneously.\\
\indent A direct comparison with our previous Taylor-based analysis~\cite{Morais:2026qdw} highlights the complementary strengths of the two cosmographic representations. At fixed catalogue size, the third-order Taylor expansion generally yields narrower constraints on the common parameters $H_0$, $q_0$, and $j_0$, as expected from its lower-dimensional parameter space. For example, at $N=100$, ET yields relative uncertainties of approximately $1.6\%$, $39.7\%$, and $135.9\%$ on $H_0$, $q_0$, and $j_0$, respectively, with the Taylor expansion, compared with approximately $2.5\%$, $40.7\%$, and $144\%$ with $P_{3,2}$. For DECIGO at the same catalogue size, the corresponding values change from $1.2\%$, $27.9\%$, and $80.6\%$ to approximately $1.7\%$, $34.1\%$, and $140\%$. This difference reflects, at least in part, the additional freedom introduced by simultaneously fitting $s_0$ and $l_0$ in the five-parameter Padé description. The comparison changes, however, when the larger event population enabled by the extended Padé redshift range is taken into account. For ET, increasing the reference catalogue from $N_z=100$ in the Taylor analysis to $N_z=470$ with $P_{3,2}$ reduces the relative uncertainty on $H_0$ from approximately $1.6\%$ to $1.4\%$ and that on $q_0$ from $39.7\%$ to approximately $31.3\%$, while the constraint on $j_0$ remains at a comparable level, approximately $136\%$ and $135\%$, respectively. Thus, for the lower-order parameters, the additional events available at $0.7\lesssim z\lesssim1.6$ can compensate for the enlarged parameter space and, in the ET case, lead to tighter constraints than those obtained with the low-redshift Taylor catalogue. For DECIGO, the increase from $N_z=800$ to $N_z=4700$ preserves a subpercent determination of $H_0$, with uncertainties of approximately $0.4\%$ and $0.5\%$ in the Taylor and Padé analyses, respectively, although the constraints on $q_0$ and especially $j_0$ remain broader in the five-parameter case.\\
\indent Beyond numerical precision alone, the Padé parametrization also provides two capabilities absent from the third-order Taylor analysis: it extends the model-independent reconstruction to substantially higher redshift, making the LISA bright-siren population accessible to the analysis, and it allows the higher-order cosmographic parameters $s_0$ and $l_0$ to be simultaneously investigated. The comparison therefore reveals a trade-off rather than a uniformly superior parametrization: the Taylor expansion provides tighter constraints within its restricted low-redshift and lower-dimensional domain, whereas $P_{3,2}$ sacrifices part of this statistical precision in exchange for a broader redshift reach, larger usable catalogues, and access to a higher-order description of the expansion history.\\
\indent Taken together, these results show that Padé cosmography provides a practical route for extending model-independent gravitational-wave studies of the expansion history beyond the conventional low-redshift Taylor regime. The broader redshift coverage enables the use of a significantly larger fraction of the populations expected for future ground- and space-based observatories, while the five-parameter analysis makes explicit how the constraining power changes as progressively higher derivatives of the scale factor are introduced. Although gravitational-wave standard sirens provide useful constraints on the lower-order parameters $H_0$ and $q_0$, the uncertainty on $j_0$ remains too large to distinguish the flat $\Lambda$CDM scenario, characterized by $j_0=1$, from models with dynamical dark energy using gravitational-wave observations alone. The situation becomes even more restrictive for the higher-order parameters $s_0$ and $l_0$. This limitation is common to both the Taylor analysis of Paper I and the Padé analysis presented here, indicating that such a discrimination will likely require complementary cosmological probes. Bright standard sirens from LISA, ET, and DECIGO can nevertheless probe complementary regions of the cosmographic expansion and provide an independent contribution to future reconstructions of the late-time Universe.

\section*{Acknowledgments}

JM acknowledges support from Conselho Nacional de Desenvolvimento Científico e Tecnológico (CNPq). JSA is supported by CNPq grants No. 307683/2022-2 and 448158/2025-6, and from FAPERJ under grant No. 299312 (2023). The development of this work was aided by the National Observatory Data Center (CPDON).

\bibliographystyle{elsarticle-num} 
\bibliography{references}

\clearpage
\onecolumn
\appendix
\section{Padé Approximants Equations}
\label{app:pade_expressions}
For completeness, we present here the explicit Padé parametrizations used in our analysis. Following the Padé construction discussed in Sec.~\ref{sec:methodology} and Ref.~\cite{Aviles:2014pade}, the approximants are obtained by matching the rational functions to the corresponding cosmographic expansion of the luminosity distance around $z=0$. We adopt natural units with $c=1$, so that the luminosity distance carries an overall factor $H_0^{-1}$.\\
\indent The $P_{2,1}$ approximant, which depends on $H_0$, $q_0$, and $j_0$, is given by
\begin{equation} \label{p21}
P_{2,1}(z) = \frac{1}{H_0} \left[
\frac{z\left(6(-1 + q_0) + (-5 - 2j_0 + q_0(8 + 3q_0))z\right)}
{-2(3 + z + j_0 z) + 2q_0(3 + z + 3q_0 z)}
\right],
\end{equation}
\indent The $P_{2,2}$ approximant additionally introduces the snap parameter $s_0$ and reads
\begin{align} \label{p22}
P_{2,2}(z) &= \frac{1}{H_0} \left(
6z(10 + 9z - 6q_0^3 z + s_0 z - 2q_0^2(3 + 7z) - q_0(16 + 19z) + j_0(4 + (9 + 6q_0)z))\Big/ \right. \nonumber \\
&\left.  (60 + 24z + 6s_0 z - 2z^2 + 4j_0^2 z^2 - 9q_0^4 z^2 - 3s_0 z^2 + 6q_0^3 z(-9 + 4z) \right. \nonumber \\
&\left. \quad + q_0^2(-36 - 114z + 19z^2)
+ j_0(24 + 6(7 + 8q_0)z + (-7 - 23q_0 + 6q_0^2)z^2) \right. \nonumber \\
&\left. \quad + q_0(-96 - 36z + (4 + 3s_0)z^2) )
\right).
\end{align}
\indent Finally, the $P_{3,2}$ approximant depends on the complete set $\{H_0,q_0,j_0,s_0,l_0\}$ considered in this work and is given by
\begin{equation}
\begin{aligned} \label{p32}
P_{3,2}(z) = \frac{1}{H_0} \Bigg(
& z\Big(-120 - 180s_0 - 156z - 36l_0 z - 426s_0 z - 40z^2 + 80j_0^3 z^2 - 30l_0 z^2 - 135q_0^6 z^2 \\
& - 210s_0 z^2 + 15s_0^2 z^2 - 270q_0^5 z(3 + 4z) + 9q_0^4(-60 + 50z + 63z^2) \\
& + 2q_0^3(720 + 1767z + 887z^2) + 3j_0^2\big(80 + 20(13 + 2q_0)z + (177 + 40q_0 - 60q_0^2)z^2\big) \\
& + 6q_0^2\big(190 + 5(67 + 9s_0)z + (125 + 3l_0 + 58s_0)z^2\big) \\
& - 6q_0\big(s_0(-30 + 4z + 17z^2) - 2(20 + (31 + 3l_0)z + (9 + 4l_0)z^2)\big) \\
& + 6j_0\big(-70 + (-127 + 10s_0)z + 45q_0^4 z^2 + (-47 - 2l_0 + 13s_0)z^2 \\
& \qquad + 5q_0^3 z(30 + 41z) - 3q_0^2(-20 + 75z + 69z^2) \\
&+ 2q_0(-115 - 274z + (-136 + 5s_0)z^2)\big)\Big) \Big/ \\
&\Big(3\big(-40 - 60s_0 - 32z - 12l_0 z - 112s_0 z - 4z^2 + 40j_0^3 z^2 - 4l_0 z^2 - 135q_0^6 z^2 \\
& - 24s_0 z^2 + 5s_0^2 z^2 - 30q_0^5 z(12 + 5z) + 3q_0^4(-60 + 160z + 71z^2) \\
& + j_0^2\big(80 + 20(11 + 4q_0)z + (57 + 20q_0 - 40q_0^2)z^2\big) \\
& + 6q_0^3(80 + 188z + (44 + 5s_0)z^2) \\
& + 2q_0^2\big(190 + 20(13 + 3s_0)z + (46 + 6l_0 + 21s_0)z^2\big) \\
& + 4q_0\big(20 + (16 + 3l_0)z + (2 + l_0)z^2 + s_0(15 - 17z - 9z^2)\big) \\
& + 2j_0\big(-70 + 2(-46 + 5s_0)z + 90q_0^4 z^2 + (-16 - 2l_0 + 3s_0)z^2 \\
& \qquad + 15q_0^3 z(12 + 5z) + q_0^2(60 - 370z - 141z^2) \\ 
&+ 2q_0(-115 - 234z + 2(-26 + 5s_0)z^2)\big)\big)\Big)
\Bigg).
\end{aligned}
\end{equation}

\section{Numerical Results for the N-based Analysis}
\label{appendix:N}


\begin{table}[H]
\centering
\scriptsize
\setlength{\tabcolsep}{3pt}
\renewcommand{\arraystretch}{1.4}

\begin{adjustbox}{max width=\textwidth}
\begin{tabular}{|c|c|c|c|c|c|}
\hline
$N$
& {$H_0$ $[\mathrm{km\,s^{-1}\,Mpc^{-1}}]$}
& {$q_0$}
& {$j_0$}
& {$s_0$}
& {$l_0$} \\
\hline

10 & $67.44^{+2.00}_{-2.11}$ & $-0.512^{+0.252}_{-0.217}$ & $1.006^{+1.350}_{-1.495}$ & $1.139^{+6.871}_{-7.440}$ & $-5.914^{+21.673}_{-17.031}$ \\ \hline
15 & $67.78^{+1.08}_{-1.19}$ & $-0.555^{+0.195}_{-0.152}$ & $1.224^{+1.118}_{-1.544}$ & $1.374^{+6.579}_{-7.757}$ & $-6.787^{+21.442}_{-16.386}$ \\ \hline
20 & $67.79^{+1.04}_{-1.23}$ & $-0.566^{+0.201}_{-0.152}$ & $1.297^{+1.136}_{-1.669}$ & $1.408^{+6.765}_{-8.268}$ & $-7.431^{+20.902}_{-15.880}$ \\ \hline
25 & $67.58^{+0.85}_{-0.94}$ & $-0.547^{+0.180}_{-0.145}$ & $1.274^{+1.073}_{-1.589}$ & $1.515^{+6.550}_{-7.939}$ & $-7.249^{+20.617}_{-15.964}$ \\ \hline
30 & $67.82^{+0.71}_{-0.83}$ & $-0.574^{+0.172}_{-0.124}$ & $1.451^{+0.928}_{-1.652}$ & $2.148^{+6.345}_{-8.236}$ & $-8.241^{+20.742}_{-15.272}$ \\ \hline
35 & $67.66^{+0.65}_{-0.78}$ & $-0.550^{+0.160}_{-0.122}$ & $1.323^{+0.925}_{-1.461}$ & $1.777^{+5.961}_{-7.379}$ & $-8.387^{+20.615}_{-15.158}$ \\ \hline
40 & $67.72^{+0.52}_{-0.63}$ & $-0.558^{+0.151}_{-0.107}$ & $1.344^{+0.861}_{-1.514}$ & $1.706^{+5.953}_{-7.906}$ & $-8.957^{+19.929}_{-14.809}$ \\ \hline

\end{tabular}
\end{adjustbox}

\caption{Cosmographic parameter estimates for LISA as a function of the number of detected events.}
\label{tab:LISA_N}
\end{table}

\begin{table}[H]
\centering
\scriptsize
\setlength{\tabcolsep}{3pt}
\renewcommand{\arraystretch}{1.4}

\begin{adjustbox}{max width=\textwidth}
\begin{tabular}{|c|c|c|c|c|c|}
\hline
$N$
& {$H_0$ $[\mathrm{km\,s^{-1}\,Mpc^{-1}}]$}
& {$q_0$}
& {$j_0$}
& {$s_0$}
& {$l_0$} \\
\hline

10   & $67.30^{+4.86}_{-4.44}$ & $-0.476^{+0.306}_{-0.290}$ & $0.921^{+1.406}_{-1.300}$ & $0.104^{+7.120}_{-8.050}$ & $-3.481^{+21.366}_{-18.331}$ \\ \hline
15   & $65.94^{+4.37}_{-3.80}$ & $-0.480^{+0.288}_{-0.270}$ & $1.027^{+1.346}_{-1.339}$ & $-1.085^{+7.349}_{-7.661}$ & $-3.659^{+20.796}_{-18.150}$ \\ \hline
20   & $66.58^{+3.55}_{-3.20}$ & $-0.485^{+0.295}_{-0.260}$ & $0.917^{+1.439}_{-1.307}$ & $-0.324^{+6.973}_{-7.753}$ & $-3.744^{+20.995}_{-18.259}$ \\ \hline
25   & $67.51^{+3.11}_{-2.89}$ & $-0.501^{+0.292}_{-0.256}$ & $0.772^{+1.527}_{-1.247}$ & $0.382^{+6.851}_{-7.742}$ & $-3.725^{+20.988}_{-18.312}$ \\ \hline
30   & $67.27^{+3.34}_{-3.03}$ & $-0.509^{+0.295}_{-0.245}$ & $0.846^{+1.487}_{-1.310}$ & $0.109^{+6.904}_{-7.838}$ & $-4.552^{+21.367}_{-17.814}$ \\ \hline
35   & $66.58^{+2.98}_{-2.73}$ & $-0.479^{+0.277}_{-0.249}$ & $0.865^{+1.473}_{-1.306}$ & $0.064^{+6.740}_{-7.570}$ & $-4.424^{+21.167}_{-17.920}$ \\ \hline
40   & $67.54^{+2.68}_{-2.65}$ & $-0.528^{+0.277}_{-0.233}$ & $0.838^{+1.495}_{-1.323}$ & $-0.070^{+6.950}_{-7.616}$ & $-4.655^{+21.241}_{-17.646}$ \\ \hline
50   & $67.73^{+2.47}_{-2.46}$ & $-0.545^{+0.283}_{-0.215}$ & $0.882^{+1.496}_{-1.378}$ & $0.347^{+6.956}_{-7.740}$ & $-5.171^{+21.562}_{-17.389}$ \\ \hline
60   & $67.42^{+2.42}_{-2.38}$ & $-0.528^{+0.279}_{-0.219}$ & $0.922^{+1.495}_{-1.408}$ & $0.371^{+6.924}_{-7.848}$ & $-4.705^{+21.258}_{-17.742}$ \\ \hline
70   & $67.65^{+2.15}_{-2.12}$ & $-0.532^{+0.247}_{-0.214}$ & $0.881^{+1.466}_{-1.381}$ & $-0.150^{+6.960}_{-7.573}$ & $-4.816^{+20.955}_{-17.539}$ \\ \hline
80   & $68.05^{+1.93}_{-1.93}$ & $-0.551^{+0.236}_{-0.205}$ & $0.851^{+1.499}_{-1.378}$ & $-0.243^{+7.067}_{-7.495}$ & $-5.163^{+20.952}_{-17.421}$ \\ \hline
90   & $67.58^{+1.84}_{-1.87}$ & $-0.534^{+0.240}_{-0.205}$ & $0.954^{+1.434}_{-1.441}$ & $0.244^{+6.699}_{-7.661}$ & $-5.282^{+20.856}_{-17.178}$ \\ \hline
100  & $67.73^{+1.65}_{-1.74}$ & $-0.535^{+0.234}_{-0.200}$ & $0.916^{+1.444}_{-1.433}$ & $0.179^{+6.893}_{-7.504}$ & $-5.540^{+21.460}_{-17.230}$ \\ \hline
150  & $67.63^{+1.54}_{-1.60}$ & $-0.545^{+0.217}_{-0.187}$ & $1.016^{+1.384}_{-1.521}$ & $0.000^{+7.075}_{-7.820}$ & $-6.247^{+21.205}_{-16.660}$ \\ \hline
200  & $67.48^{+1.41}_{-1.44}$ & $-0.500^{+0.201}_{-0.194}$ & $0.899^{+1.344}_{-1.408}$ & $0.374^{+6.663}_{-7.159}$ & $-5.129^{+21.278}_{-17.456}$ \\ \hline
250  & $67.96^{+1.11}_{-1.20}$ & $-0.569^{+0.196}_{-0.165}$ & $1.221^{+1.236}_{-1.635}$ & $0.982^{+6.894}_{-7.971}$ & $-7.084^{+21.510}_{-16.098}$ \\ \hline
300  & $67.69^{+1.00}_{-1.10}$ & $-0.541^{+0.188}_{-0.163}$ & $1.164^{+1.209}_{-1.601}$ & $0.994^{+6.784}_{-7.897}$ & $-7.119^{+21.546}_{-16.198}$ \\ \hline
350  & $67.56^{+1.06}_{-1.15}$ & $-0.522^{+0.189}_{-0.170}$ & $1.106^{+1.209}_{-1.548}$ & $1.069^{+6.722}_{-7.485}$ & $-6.690^{+21.455}_{-16.490}$ \\ \hline
400  & $67.80^{+0.95}_{-1.04}$ & $-0.556^{+0.187}_{-0.160}$ & $1.227^{+1.209}_{-1.702}$ & $0.947^{+6.940}_{-8.251}$ & $-7.449^{+21.297}_{-15.937}$ \\ \hline
450  & $67.66^{+0.94}_{-1.01}$ & $-0.543^{+0.185}_{-0.156}$ & $1.215^{+1.140}_{-1.611}$ & $1.271^{+6.562}_{-7.757}$ & $-7.287^{+21.322}_{-16.042}$ \\ \hline
470  & $67.55^{+0.92}_{-0.98}$ & $-0.532^{+0.179}_{-0.155}$ & $1.181^{+1.123}_{-1.575}$ & $1.260^{+6.536}_{-7.788}$ & $-7.094^{+21.165}_{-16.044}$ \\ \hline
500  & $67.59^{+0.83}_{-0.88}$ & $-0.525^{+0.166}_{-0.155}$ & $1.088^{+1.169}_{-1.541}$ & $0.536^{+6.618}_{-7.590}$ & $-7.579^{+21.084}_{-15.719}$ \\ \hline
600  & $67.68^{+0.77}_{-0.81}$ & $-0.543^{+0.165}_{-0.138}$ & $1.253^{+1.008}_{-1.511}$ & $1.666^{+6.039}_{-7.632}$ & $-7.813^{+21.238}_{-15.620}$ \\ \hline
700  & $67.72^{+0.73}_{-0.77}$ & $-0.556^{+0.166}_{-0.131}$ & $1.358^{+0.980}_{-1.640}$ & $1.737^{+6.324}_{-8.184}$ & $-7.807^{+21.318}_{-15.630}$ \\ \hline
800  & $67.82^{+0.67}_{-0.71}$ & $-0.568^{+0.159}_{-0.123}$ & $1.401^{+0.955}_{-1.592}$ & $1.923^{+6.019}_{-8.134}$ & $-7.859^{+21.151}_{-15.574}$ \\ \hline
900  & $67.72^{+0.60}_{-0.64}$ & $-0.561^{+0.155}_{-0.120}$ & $1.379^{+0.952}_{-1.598}$ & $1.885^{+6.203}_{-8.228}$ & $-7.859^{+20.788}_{-15.528}$ \\ \hline
1000 & $67.71^{+0.60}_{-0.66}$ & $-0.559^{+0.152}_{-0.121}$ & $1.337^{+0.957}_{-1.532}$ & $1.588^{+6.233}_{-7.863}$ & $-8.463^{+20.533}_{-15.139}$ \\ \hline

\end{tabular}
\end{adjustbox}

\caption{Cosmographic parameter estimates for the Einstein Telescope as a function of the number of detected events.}
\label{tab:ET_N}
\end{table}

\begin{table}[H]
\centering
\scriptsize
\setlength{\tabcolsep}{3pt}
\renewcommand{\arraystretch}{1.4}

\begin{adjustbox}{max width=\textwidth}
\begin{tabular}{|c|c|c|c|c|c|}
\hline
$N$
& {$H_0$ $[\mathrm{km\,s^{-1}\,Mpc^{-1}}]$}
& {$q_0$}
& {$j_0$}
& {$s_0$}
& {$l_0$} \\
\hline

10    & $67.24^{+3.50}_{-3.17}$ & $-0.504^{+0.301}_{-0.258}$ & $0.871^{+1.443}_{-1.349}$ & $0.639^{+6.828}_{-7.717}$ & $-3.815^{+21.341}_{-18.278}$ \\ \hline
15    & $67.29^{+3.08}_{-2.92}$ & $-0.535^{+0.291}_{-0.223}$ & $0.930^{+1.462}_{-1.404}$ & $-0.165^{+7.334}_{-7.911}$ & $-5.452^{+21.330}_{-17.257}$ \\ \hline
20    & $67.27^{+2.35}_{-2.35}$ & $-0.505^{+0.274}_{-0.224}$ & $0.896^{+1.470}_{-1.405}$ & $0.515^{+6.822}_{-7.587}$ & $-5.261^{+21.651}_{-17.400}$ \\ \hline
25    & $67.28^{+2.30}_{-2.41}$ & $-0.516^{+0.277}_{-0.224}$ & $0.942^{+1.456}_{-1.442}$ & $0.883^{+6.992}_{-7.617}$ & $-5.141^{+21.537}_{-17.527}$ \\ \hline
30    & $67.79^{+1.97}_{-2.11}$ & $-0.541^{+0.249}_{-0.207}$ & $0.935^{+1.436}_{-1.448}$ & $0.171^{+6.948}_{-7.593}$ & $-5.299^{+21.179}_{-17.360}$ \\ \hline
35    & $67.46^{+1.70}_{-1.74}$ & $-0.531^{+0.227}_{-0.198}$ & $1.001^{+1.370}_{-1.508}$ & $0.147^{+7.066}_{-7.689}$ & $-5.268^{+21.219}_{-17.294}$ \\ \hline
40    & $67.54^{+1.77}_{-1.83}$ & $-0.529^{+0.229}_{-0.203}$ & $1.019^{+1.344}_{-1.485}$ & $0.671^{+6.856}_{-7.602}$ & $-5.247^{+20.950}_{-17.350}$ \\ \hline
50    & $67.81^{+1.76}_{-1.84}$ & $-0.539^{+0.228}_{-0.207}$ & $1.020^{+1.397}_{-1.522}$ & $0.305^{+6.951}_{-7.446}$ & $-5.670^{+21.105}_{-16.982}$ \\ \hline
70    & $67.61^{+1.33}_{-1.52}$ & $-0.541^{+0.223}_{-0.184}$ & $1.166^{+1.304}_{-1.616}$ & $1.013^{+6.851}_{-7.727}$ & $-6.767^{+21.275}_{-16.359}$ \\ \hline
90    & $67.61^{+1.19}_{-1.30}$ & $-0.535^{+0.201}_{-0.176}$ & $1.196^{+1.192}_{-1.604}$ & $1.307^{+6.512}_{-7.726}$ & $-6.799^{+21.629}_{-16.460}$ \\ \hline
100   & $67.70^{+1.13}_{-1.21}$ & $-0.543^{+0.197}_{-0.167}$ & $1.179^{+1.192}_{-1.605}$ & $1.073^{+6.882}_{-7.708}$ & $-6.879^{+21.280}_{-16.280}$ \\ \hline
200   & $67.86^{+0.81}_{-0.96}$ & $-0.581^{+0.184}_{-0.138}$ & $1.465^{+1.044}_{-1.698}$ & $2.070^{+6.657}_{-8.368}$ & $-8.191^{+21.508}_{-15.351}$ \\ \hline
300   & $67.81^{+0.75}_{-0.84}$ & $-0.563^{+0.170}_{-0.125}$ & $1.389^{+0.916}_{-1.603}$ & $2.004^{+5.965}_{-7.922}$ & $-7.866^{+21.381}_{-15.498}$ \\ \hline
400   & $67.81^{+0.69}_{-0.80}$ & $-0.577^{+0.175}_{-0.118}$ & $1.500^{+0.883}_{-1.666}$ & $2.421^{+5.836}_{-8.355}$ & $-8.590^{+20.546}_{-15.033}$ \\ \hline
500   & $67.88^{+0.60}_{-0.70}$ & $-0.590^{+0.158}_{-0.106}$ & $1.579^{+0.829}_{-1.526}$ & $2.831^{+5.713}_{-7.989}$ & $-8.426^{+20.449}_{-15.096}$ \\ \hline
700   & $67.76^{+0.60}_{-0.66}$ & $-0.565^{+0.151}_{-0.112}$ & $1.387^{+0.899}_{-1.454}$ & $1.959^{+5.927}_{-7.581}$ & $-8.519^{+20.195}_{-14.907}$ \\ \hline
1000  & $67.76^{+0.52}_{-0.60}$ & $-0.570^{+0.145}_{-0.096}$ & $1.471^{+0.746}_{-1.413}$ & $2.503^{+5.162}_{-7.628}$ & $-8.862^{+19.193}_{-14.633}$ \\ \hline
1500  & $67.86^{+0.44}_{-0.57}$ & $-0.591^{+0.137}_{-0.084}$ & $1.614^{+0.685}_{-1.338}$ & $3.214^{+5.020}_{-7.561}$ & $-8.778^{+19.522}_{-14.800}$ \\ \hline
2000  & $67.86^{+0.41}_{-0.48}$ & $-0.588^{+0.116}_{-0.081}$ & $1.552^{+0.713}_{-1.142}$ & $2.700^{+5.178}_{-6.577}$ & $-9.298^{+19.113}_{-14.482}$ \\ \hline
2500  & $67.81^{+0.38}_{-0.48}$ & $-0.582^{+0.118}_{-0.075}$ & $1.538^{+0.646}_{-1.172}$ & $2.638^{+4.743}_{-6.733}$ & $-9.777^{+18.906}_{-14.061}$ \\ \hline
3000  & $67.77^{+0.37}_{-0.42}$ & $-0.569^{+0.102}_{-0.076}$ & $1.391^{+0.682}_{-1.028}$ & $1.972^{+4.749}_{-6.288}$ & $-9.493^{+18.473}_{-14.253}$ \\ \hline
3500  & $67.82^{+0.36}_{-0.41}$ & $-0.575^{+0.103}_{-0.076}$ & $1.429^{+0.697}_{-1.041}$ & $1.965^{+4.864}_{-6.185}$ & $-10.131^{+18.849}_{-13.950}$ \\ \hline
4000  & $67.84^{+0.31}_{-0.40}$ & $-0.587^{+0.100}_{-0.061}$ & $1.569^{+0.543}_{-1.010}$ & $2.755^{+4.118}_{-6.097}$ & $-10.517^{+18.819}_{-13.687}$ \\ \hline
4500  & $67.83^{+0.29}_{-0.38}$ & $-0.584^{+0.094}_{-0.060}$ & $1.553^{+0.531}_{-0.970}$ & $2.681^{+4.040}_{-6.017}$ & $-10.810^{+19.341}_{-13.498}$ \\ \hline
4700  & $67.75^{+0.31}_{-0.35}$ & $-0.562^{+0.086}_{-0.069}$ & $1.325^{+0.639}_{-0.874}$ & $1.515^{+4.293}_{-5.458}$ & $-10.366^{+18.407}_{-13.820}$ \\ \hline
5000  & $67.75^{+0.31}_{-0.36}$ & $-0.566^{+0.091}_{-0.067}$ & $1.373^{+0.613}_{-0.943}$ & $1.729^{+4.372}_{-5.813}$ & $-10.642^{+18.424}_{-13.582}$ \\ \hline
6000  & $67.77^{+0.29}_{-0.36}$ & $-0.568^{+0.092}_{-0.064}$ & $1.371^{+0.608}_{-0.945}$ & $1.431^{+4.357}_{-5.644}$ & $-11.516^{+18.396}_{-12.946}$ \\ \hline
7000  & $67.75^{+0.29}_{-0.34}$ & $-0.561^{+0.089}_{-0.067}$ & $1.292^{+0.647}_{-0.921}$ & $1.064^{+4.487}_{-5.559}$ & $-11.429^{+18.638}_{-13.048}$ \\ \hline
8000  & $67.79^{+0.23}_{-0.30}$ & $-0.574^{+0.077}_{-0.050}$ & $1.451^{+0.464}_{-0.806}$ & $1.985^{+3.558}_{-5.164}$ & $-10.719^{+18.623}_{-13.472}$ \\ \hline
9000  & $67.77^{+0.25}_{-0.30}$ & $-0.565^{+0.078}_{-0.057}$ & $1.326^{+0.549}_{-0.818}$ & $1.177^{+4.001}_{-5.107}$ & $-11.150^{+18.237}_{-13.254}$ \\ \hline
10000 & $67.82^{+0.23}_{-0.31}$ & $-0.579^{+0.083}_{-0.051}$ & $1.466^{+0.506}_{-0.872}$ & $1.675^{+3.955}_{-5.229}$ & $-11.890^{+18.899}_{-12.795}$ \\ \hline

\end{tabular}
\end{adjustbox}

\caption{Cosmographic parameter estimates for DECIGO as a function of the number of detected events.}
\label{tab:DECIGO_N}
\end{table}

\clearpage

\section{Numerical Results for the SNR-based Analysis}
\label{appendix:SNR}

\vspace{1em}

\begin{table}[H]
\centering
\scriptsize
\setlength{\tabcolsep}{3pt}
\renewcommand{\arraystretch}{1.4}

\begin{adjustbox}{max width=\textwidth}
\begin{tabular}{|c|c|c|c|c|c|c|}
\hline
SNR interval
& $\bar{z}$
& {$H_0$ $[\mathrm{km\,s^{-1}\,Mpc^{-1}}]$}
& {$q_0$}
& {$j_0$}
& {$s_0$}
& {$l_0$} \\
\hline

$[8,50]$       & 1.414 & $65.07^{+6.50}_{-5.34}$ & $-0.468^{+0.290}_{-0.284}$ & $1.148^{+1.227}_{-1.484}$ & $0.099^{+7.700}_{-7.559}$ & $-5.389^{+21.818}_{-17.372}$ \\ \hline
$[50,100]$     & 1.357 & $66.47^{+5.30}_{-4.54}$ & $-0.507^{+0.320}_{-0.254}$ & $1.092^{+1.291}_{-1.510}$ & $1.084^{+7.456}_{-7.832}$ & $-6.076^{+22.197}_{-16.986}$ \\ \hline
$[100,150]$    & 1.321 & $67.04^{+4.81}_{-4.39}$ & $-0.536^{+0.326}_{-0.233}$ & $1.076^{+1.297}_{-1.513}$ & $0.969^{+7.678}_{-7.941}$ & $-5.969^{+21.859}_{-16.954}$ \\ \hline
$[150,200]$    & 1.298 & $66.62^{+4.46}_{-4.01}$ & $-0.516^{+0.320}_{-0.239}$ & $1.050^{+1.330}_{-1.490}$ & $0.760^{+7.604}_{-7.725}$ & $-5.920^{+21.776}_{-16.916}$ \\ \hline
$[200,300]$    & 1.243 & $66.81^{+3.86}_{-3.56}$ & $-0.504^{+0.314}_{-0.245}$ & $0.954^{+1.405}_{-1.422}$ & $0.850^{+7.095}_{-7.582}$ & $-5.674^{+21.915}_{-17.269}$ \\ \hline
$[300,400]$    & 1.224 & $67.75^{+3.71}_{-3.64}$ & $-0.535^{+0.318}_{-0.231}$ & $0.941^{+1.399}_{-1.414}$ & $0.922^{+7.161}_{-7.526}$ & $-5.698^{+21.704}_{-17.189}$ \\ \hline
$[400,500]$    & 1.212 & $67.31^{+3.68}_{-3.48}$ & $-0.521^{+0.313}_{-0.238}$ & $0.958^{+1.375}_{-1.425}$ & $0.895^{+7.131}_{-7.412}$ & $-5.237^{+21.691}_{-17.444}$ \\ \hline
$[500,600]$    & 1.206 & $67.25^{+3.56}_{-3.42}$ & $-0.522^{+0.307}_{-0.236}$ & $0.948^{+1.408}_{-1.418}$ & $0.598^{+7.152}_{-7.390}$ & $-5.269^{+21.625}_{-17.393}$ \\ \hline
$[600,700]$    & 1.201 & $67.40^{+3.39}_{-3.35}$ & $-0.531^{+0.317}_{-0.226}$ & $0.956^{+1.419}_{-1.429}$ & $0.796^{+7.267}_{-7.514}$ & $-6.053^{+22.062}_{-16.905}$ \\ \hline
$[700,800]$    & 1.190 & $67.43^{+3.27}_{-3.24}$ & $-0.541^{+0.298}_{-0.219}$ & $0.946^{+1.422}_{-1.434}$ & $0.336^{+7.428}_{-7.590}$ & $-6.118^{+21.530}_{-16.810}$ \\ \hline
$[800,900]$    & 1.183 & $67.27^{+3.18}_{-3.14}$ & $-0.521^{+0.300}_{-0.226}$ & $0.942^{+1.438}_{-1.434}$ & $0.539^{+7.278}_{-7.500}$ & $-5.548^{+21.429}_{-17.228}$ \\ \hline
$[900,1000]$   & 1.160 & $67.42^{+3.00}_{-3.08}$ & $-0.532^{+0.300}_{-0.220}$ & $0.892^{+1.500}_{-1.481}$ & $0.245^{+7.356}_{-7.494}$ & $-5.952^{+21.453}_{-16.907}$ \\ \hline
$[1000,1500]$  & 1.081 & $67.48^{+2.43}_{-2.62}$ & $-0.514^{+0.276}_{-0.229}$ & $0.887^{+1.487}_{-1.413}$ & $0.538^{+7.077}_{-7.335}$ & $-5.588^{+21.506}_{-17.201}$ \\ \hline
$[1500,2000]$  & 0.943 & $67.05^{+2.26}_{-2.36}$ & $-0.482^{+0.259}_{-0.237}$ & $0.825^{+1.496}_{-1.322}$ & $0.219^{+6.522}_{-7.100}$ & $-4.224^{+20.909}_{-18.016}$ \\ \hline
$[2000,2500]$  & 0.854 & $67.58^{+2.13}_{-2.28}$ & $-0.517^{+0.260}_{-0.220}$ & $0.787^{+1.530}_{-1.316}$ & $-0.082^{+6.779}_{-7.474}$ & $-4.196^{+20.805}_{-17.941}$ \\ \hline
$[2500,3000]$  & 0.800 & $67.40^{+2.11}_{-2.15}$ & $-0.508^{+0.250}_{-0.223}$ & $0.766^{+1.513}_{-1.276}$ & $-0.314^{+6.691}_{-7.333}$ & $-3.523^{+20.617}_{-18.343}$ \\ \hline
$[3000,3500]$  & 0.749 & $67.47^{+2.09}_{-2.21}$ & $-0.499^{+0.258}_{-0.239}$ & $0.702^{+1.556}_{-1.227}$ & $0.244^{+6.492}_{-7.526}$ & $-3.113^{+20.725}_{-18.579}$ \\ \hline
$[3500,4000]$  & 0.707 & $67.57^{+1.83}_{-1.96}$ & $-0.514^{+0.248}_{-0.217}$ & $0.750^{+1.525}_{-1.261}$ & $-0.069^{+6.626}_{-7.683}$ & $-3.111^{+20.573}_{-18.489}$ \\ \hline
$[4000,4500]$  & 0.679 & $67.63^{+1.83}_{-2.06}$ & $-0.516^{+0.259}_{-0.224}$ & $0.714^{+1.564}_{-1.214}$ & $-0.039^{+6.535}_{-7.640}$ & $-2.780^{+20.479}_{-18.766}$ \\ \hline
$[4500,5000]$  & 0.655 & $67.47^{+1.80}_{-1.96}$ & $-0.500^{+0.247}_{-0.222}$ & $0.683^{+1.558}_{-1.171}$ & $-0.116^{+6.565}_{-7.502}$ & $-2.303^{+20.290}_{-18.728}$ \\ \hline
$[5000,5500]$  & 0.627 & $67.81^{+1.85}_{-1.96}$ & $-0.526^{+0.251}_{-0.227}$ & $0.700^{+1.544}_{-1.184}$ & $-0.341^{+6.687}_{-7.782}$ & $-2.708^{+20.330}_{-18.655}$ \\ \hline
$[5500,6000]$  & 0.606 & $67.38^{+1.82}_{-1.88}$ & $-0.480^{+0.242}_{-0.230}$ & $0.675^{+1.445}_{-1.148}$ & $0.462^{+6.507}_{-7.673}$ & $-1.526^{+19.798}_{-19.048}$ \\ \hline
$[6000,6500]$  & 0.589 & $67.41^{+1.71}_{-1.80}$ & $-0.487^{+0.239}_{-0.233}$ & $0.705^{+1.487}_{-1.167}$ & $0.318^{+6.541}_{-7.777}$ & $-2.013^{+20.136}_{-19.017}$ \\ \hline
$[6500,7000]$  & 0.570 & $67.24^{+1.69}_{-1.75}$ & $-0.470^{+0.238}_{-0.228}$ & $0.670^{+1.473}_{-1.147}$ & $0.481^{+6.565}_{-7.860}$ & $-1.483^{+19.956}_{-19.183}$ \\ \hline
$[7000,7500]$  & 0.559 & $67.69^{+1.67}_{-1.88}$ & $-0.528^{+0.259}_{-0.216}$ & $0.723^{+1.524}_{-1.157}$ & $-0.345^{+6.819}_{-7.930}$ & $-1.832^{+19.897}_{-19.198}$ \\ \hline
$[7500,8000]$  & 0.545 & $67.31^{+1.62}_{-1.70}$ & $-0.476^{+0.235}_{-0.225}$ & $0.682^{+1.466}_{-1.133}$ & $0.589^{+6.634}_{-7.908}$ & $-1.342^{+19.762}_{-19.261}$ \\ \hline
$[8000,8500]$  & 0.533 & $67.77^{+1.56}_{-1.76}$ & $-0.538^{+0.250}_{-0.211}$ & $0.749^{+1.519}_{-1.166}$ & $-0.500^{+7.032}_{-7.974}$ & $-2.072^{+19.958}_{-19.009}$ \\ \hline
$[8500,9000]$  & 0.521 & $67.37^{+1.64}_{-1.59}$ & $-0.475^{+0.222}_{-0.229}$ & $0.675^{+1.381}_{-1.094}$ & $0.298^{+6.579}_{-7.736}$ & $-0.769^{+19.483}_{-19.554}$ \\ \hline
$[9000,9500]$  & 0.511 & $67.36^{+1.64}_{-1.70}$ & $-0.476^{+0.241}_{-0.236}$ & $0.683^{+1.444}_{-1.118}$ & $0.725^{+6.782}_{-8.194}$ & $-0.939^{+19.700}_{-19.586}$ \\ \hline
$[9500,10000]$ & 0.502 & $67.51^{+1.65}_{-1.63}$ & $-0.497^{+0.231}_{-0.231}$ & $0.704^{+1.423}_{-1.098}$ & $0.113^{+6.895}_{-7.881}$ & $-0.850^{+19.605}_{-19.502}$ \\ \hline

\end{tabular}
\end{adjustbox}

\caption{Cosmographic parameter estimates for LISA as a function of the signal-to-noise ratio. The quantity $\bar{z}$ denotes the mean redshift associated with each SNR interval.}
\label{tab:LISA_SNR}
\end{table}

\clearpage

\begin{table}[H]
\centering
\scriptsize
\setlength{\tabcolsep}{3pt}
\renewcommand{\arraystretch}{1.4}

\begin{adjustbox}{max width=\textwidth}
\begin{tabular}{|c|c|c|c|c|c|c|}
\hline
SNR interval
& $\bar{z}$
& {$H_0$ $[\mathrm{km\,s^{-1}\,Mpc^{-1}}]$}
& {$q_0$}
& {$j_0$}
& {$s_0$}
& {$l_0$} \\
\hline

$[8,23]$    & 1.142 & $66.78^{+4.95}_{-4.34}$ & $-0.505^{+0.311}_{-0.261}$ & $0.973^{+1.390}_{-1.409}$ & $0.274^{+7.400}_{-7.831}$ & $-5.589^{+21.720}_{-17.222}$ \\ \hline
$[23,38]$   & 1.086 & $67.01^{+3.51}_{-3.25}$ & $-0.511^{+0.293}_{-0.241}$ & $0.850^{+1.496}_{-1.336}$ & $-0.125^{+7.128}_{-7.533}$ & $-5.052^{+21.103}_{-17.436}$ \\ \hline
$[38,53]$   & 0.979 & $67.69^{+2.72}_{-2.73}$ & $-0.541^{+0.280}_{-0.217}$ & $0.839^{+1.517}_{-1.353}$ & $-0.196^{+6.994}_{-7.525}$ & $-5.466^{+21.145}_{-17.130}$ \\ \hline
$[53,68]$   & 0.697 & $67.48^{+2.21}_{-2.22}$ & $-0.507^{+0.264}_{-0.238}$ & $0.686^{+1.564}_{-1.183}$ & $-0.254^{+6.598}_{-7.580}$ & $-3.015^{+20.483}_{-18.567}$ \\ \hline
$[68,83]$   & 0.533 & $67.60^{+1.79}_{-1.83}$ & $-0.507^{+0.248}_{-0.231}$ & $0.699^{+1.465}_{-1.127}$ & $-0.028^{+6.907}_{-7.969}$ & $-1.328^{+19.772}_{-19.217}$ \\ \hline
$[83,98]$   & 0.432 & $67.52^{+1.55}_{-1.54}$ & $-0.491^{+0.239}_{-0.238}$ & $0.694^{+1.397}_{-1.086}$ & $0.219^{+7.550}_{-8.270}$ & $-0.304^{+19.652}_{-19.867}$ \\ \hline
$[98,113]$  & 0.360 & $67.78^{+1.27}_{-1.30}$ & $-0.538^{+0.227}_{-0.214}$ & $0.858^{+1.305}_{-1.150}$ & $-0.475^{+8.207}_{-8.328}$ & $0.261^{+19.712}_{-20.345}$ \\ \hline
$[113,128]$ & 0.311 & $67.67^{+1.07}_{-1.04}$ & $-0.525^{+0.203}_{-0.203}$ & $0.901^{+1.252}_{-1.171}$ & $-0.370^{+8.474}_{-8.583}$ & $0.587^{+19.850}_{-20.380}$ \\ \hline
$[128,143]$ & 0.274 & $67.54^{+0.95}_{-0.92}$ & $-0.513^{+0.195}_{-0.201}$ & $0.955^{+1.299}_{-1.242}$ & $-0.879^{+9.259}_{-8.921}$ & $0.538^{+19.945}_{-20.523}$ \\ \hline
$[143,158]$ & 0.243 & $67.70^{+0.78}_{-0.76}$ & $-0.534^{+0.178}_{-0.180}$ & $1.009^{+1.259}_{-1.266}$ & $-1.025^{+9.456}_{-8.941}$ & $0.631^{+19.962}_{-20.603}$ \\ \hline
$[158,173]$ & 0.221 & $67.62^{+0.68}_{-0.65}$ & $-0.525^{+0.159}_{-0.171}$ & $0.943^{+1.297}_{-1.255}$ & $-0.247^{+9.485}_{-9.401}$ & $0.190^{+20.138}_{-20.347}$ \\ \hline
$[173,188]$ & 0.200 & $67.65^{+0.58}_{-0.56}$ & $-0.528^{+0.150}_{-0.159}$ & $0.964^{+1.302}_{-1.275}$ & $-0.488^{+9.785}_{-9.612}$ & $0.262^{+20.112}_{-20.274}$ \\ \hline
$[188,203]$ & 0.184 & $67.66^{+0.53}_{-0.51}$ & $-0.537^{+0.145}_{-0.146}$ & $1.067^{+1.268}_{-1.312}$ & $-1.062^{+9.976}_{-9.322}$ & $0.412^{+20.089}_{-20.482}$ \\ \hline
$[203,218]$ & 0.170 & $67.55^{+0.46}_{-0.45}$ & $-0.508^{+0.133}_{-0.139}$ & $0.955^{+1.318}_{-1.277}$ & $-0.221^{+9.853}_{-9.757}$ & $0.363^{+20.124}_{-20.447}$ \\ \hline
$[218,233]$ & 0.158 & $67.64^{+0.42}_{-0.42}$ & $-0.523^{+0.128}_{-0.132}$ & $0.989^{+1.306}_{-1.306}$ & $-0.422^{+10.081}_{-9.741}$ & $0.177^{+20.312}_{-20.487}$ \\ \hline
$[233,248]$ & 0.148 & $67.67^{+0.39}_{-0.38}$ & $-0.535^{+0.124}_{-0.124}$ & $1.059^{+1.290}_{-1.334}$ & $-0.780^{+10.159}_{-9.653}$ & $0.308^{+20.102}_{-20.453}$ \\ \hline
$[248,263]$ & 0.140 & $67.64^{+0.36}_{-0.35}$ & $-0.526^{+0.116}_{-0.122}$ & $0.907^{+1.347}_{-1.275}$ & $0.296^{+9.872}_{-10.122}$ & $-0.084^{+20.423}_{-20.277}$ \\ \hline
$[263,278]$ & 0.131 & $67.62^{+0.33}_{-0.33}$ & $-0.528^{+0.113}_{-0.116}$ & $1.012^{+1.323}_{-1.327}$ & $-0.374^{+10.182}_{-9.869}$ & $0.157^{+20.275}_{-20.326}$ \\ \hline
$[278,293]$ & 0.123 & $67.64^{+0.30}_{-0.30}$ & $-0.525^{+0.108}_{-0.110}$ & $0.965^{+1.335}_{-1.311}$ & $-0.168^{+10.124}_{-9.969}$ & $0.113^{+20.275}_{-20.348}$ \\ \hline
$[293,308]$ & 0.117 & $67.65^{+0.29}_{-0.29}$ & $-0.528^{+0.106}_{-0.107}$ & $0.957^{+1.337}_{-1.304}$ & $-0.064^{+10.019}_{-10.038}$ & $-0.044^{+20.236}_{-20.248}$ \\ \hline

\end{tabular}
\end{adjustbox}

\caption{Cosmographic parameter estimates for the Einstein Telescope as a function of the signal-to-noise ratio. The quantity $\bar{z}$ denotes the mean redshift associated with each SNR interval.}
\label{tab:ET_SNR}
\end{table}

\clearpage

\begin{table}[H]
\centering
\scriptsize
\setlength{\tabcolsep}{3pt}
\renewcommand{\arraystretch}{1.4}

\begin{adjustbox}{max width=\textwidth}
\begin{tabular}{|c|c|c|c|c|c|c|}
\hline
SNR interval
& $\bar{z}$
& {$H_0$ $[\mathrm{km\,s^{-1}\,Mpc^{-1}}]$}
& {$q_0$}
& {$j_0$}
& {$s_0$}
& {$l_0$} \\
\hline

$[8,48]$     & 1.320 & $66.71^{+5.19}_{-4.55}$ & $-0.522^{+0.317}_{-0.243}$ & $1.079^{+1.308}_{-1.504}$ & $0.766^{+7.579}_{-7.838}$ & $-6.454^{+22.040}_{-16.672}$ \\ \hline
$[48,88]$    & 1.133 & $67.31^{+2.94}_{-3.02}$ & $-0.513^{+0.292}_{-0.233}$ & $0.932^{+1.436}_{-1.426}$ & $0.646^{+7.016}_{-7.313}$ & $-5.730^{+21.699}_{-17.159}$ \\ \hline
$[88,128]$   & 1.092 & $67.69^{+2.17}_{-2.43}$ & $-0.529^{+0.258}_{-0.218}$ & $1.018^{+1.381}_{-1.505}$ & $0.826^{+7.004}_{-7.342}$ & $-6.276^{+21.648}_{-16.701}$ \\ \hline
$[128,168]$  & 1.051 & $67.62^{+1.68}_{-1.82}$ & $-0.522^{+0.228}_{-0.200}$ & $1.046^{+1.301}_{-1.561}$ & $0.722^{+6.815}_{-7.351}$ & $-6.550^{+21.236}_{-16.537}$ \\ \hline
$[168,208]$  & 0.891 & $67.73^{+1.36}_{-1.53}$ & $-0.540^{+0.215}_{-0.189}$ & $1.108^{+1.307}_{-1.599}$ & $0.756^{+6.781}_{-7.653}$ & $-6.399^{+21.374}_{-16.670}$ \\ \hline
$[208,248]$  & 0.709 & $67.51^{+1.08}_{-1.22}$ & $-0.513^{+0.191}_{-0.179}$ & $0.938^{+1.372}_{-1.461}$ & $0.401^{+6.747}_{-7.562}$ & $-5.264^{+20.946}_{-17.223}$ \\ \hline
$[248,288]$  & 0.580 & $67.71^{+0.95}_{-1.03}$ & $-0.520^{+0.171}_{-0.171}$ & $0.833^{+1.394}_{-1.354}$ & $-0.275^{+6.805}_{-7.545}$ & $-4.439^{+20.892}_{-17.896}$ \\ \hline
$[288,328]$  & 0.496 & $67.71^{+0.86}_{-0.91}$ & $-0.528^{+0.162}_{-0.169}$ & $0.813^{+1.444}_{-1.278}$ & $-0.729^{+6.893}_{-7.615}$ & $-3.202^{+20.361}_{-18.551}$ \\ \hline
$[328,368]$  & 0.433 & $67.65^{+0.78}_{-0.82}$ & $-0.516^{+0.156}_{-0.157}$ & $0.759^{+1.378}_{-1.200}$ & $-0.630^{+6.787}_{-7.674}$ & $-1.988^{+19.992}_{-19.055}$ \\ \hline
$[368,408]$  & 0.383 & $67.58^{+0.74}_{-0.76}$ & $-0.499^{+0.148}_{-0.157}$ & $0.642^{+1.360}_{-1.078}$ & $-0.565^{+6.890}_{-7.587}$ & $-0.666^{+19.706}_{-19.728}$ \\ \hline
$[408,448]$  & 0.345 & $67.53^{+0.69}_{-0.69}$ & $-0.495^{+0.142}_{-0.151}$ & $0.690^{+1.284}_{-1.057}$ & $-0.059^{+6.853}_{-7.623}$ & $-0.365^{+19.708}_{-19.939}$ \\ \hline
$[448,488]$  & 0.311 & $67.72^{+0.64}_{-0.64}$ & $-0.534^{+0.139}_{-0.145}$ & $0.823^{+1.265}_{-1.067}$ & $-1.218^{+7.224}_{-7.598}$ & $0.380^{+19.693}_{-20.396}$ \\ \hline
$[488,528]$  & 0.287 & $67.66^{+0.61}_{-0.64}$ & $-0.525^{+0.144}_{-0.146}$ & $0.833^{+1.248}_{-1.091}$ & $-0.726^{+7.467}_{-7.922}$ & $0.617^{+19.644}_{-20.554}$ \\ \hline
$[528,568]$  & 0.265 & $67.61^{+0.60}_{-0.58}$ & $-0.521^{+0.139}_{-0.144}$ & $0.879^{+1.177}_{-1.097}$ & $-0.510^{+7.800}_{-8.092}$ & $0.522^{+19.723}_{-20.453}$ \\ \hline
$[568,608]$  & 0.244 & $67.71^{+0.57}_{-0.58}$ & $-0.542^{+0.145}_{-0.145}$ & $0.987^{+1.188}_{-1.146}$ & $-1.369^{+8.205}_{-7.929}$ & $0.647^{+19.847}_{-20.527}$ \\ \hline
$[608,648]$  & 0.229 & $67.71^{+0.55}_{-0.52}$ & $-0.538^{+0.137}_{-0.142}$ & $0.951^{+1.172}_{-1.131}$ & $-0.802^{+8.396}_{-8.364}$ & $0.868^{+19.679}_{-20.530}$ \\ \hline
$[648,688]$  & 0.215 & $67.63^{+0.52}_{-0.49}$ & $-0.516^{+0.133}_{-0.141}$ & $0.826^{+1.216}_{-1.108}$ & $0.543^{+8.340}_{-8.933}$ & $0.318^{+20.005}_{-20.368}$ \\ \hline
$[688,728]$  & 0.202 & $67.66^{+0.50}_{-0.49}$ & $-0.537^{+0.140}_{-0.140}$ & $1.042^{+1.179}_{-1.199}$ & $-1.281^{+9.054}_{-8.625}$ & $0.726^{+19.872}_{-20.579}$ \\ \hline
$[728,768]$  & 0.191 & $67.73^{+0.48}_{-0.47}$ & $-0.550^{+0.138}_{-0.138}$ & $1.061^{+1.214}_{-1.249}$ & $-1.080^{+9.395}_{-8.796}$ & $0.593^{+19.930}_{-20.527}$ \\ \hline
$[768,808]$  & 0.180 & $67.66^{+0.45}_{-0.43}$ & $-0.528^{+0.131}_{-0.138}$ & $0.953^{+1.240}_{-1.216}$ & $-0.231^{+9.319}_{-9.361}$ & $0.318^{+20.086}_{-20.377}$ \\ \hline
$[808,848]$  & 0.171 & $67.58^{+0.44}_{-0.41}$ & $-0.509^{+0.127}_{-0.138}$ & $0.878^{+1.298}_{-1.214}$ & $0.327^{+9.353}_{-9.748}$ & $0.283^{+20.006}_{-20.346}$ \\ \hline
$[848,888]$  & 0.164 & $67.65^{+0.42}_{-0.40}$ & $-0.531^{+0.127}_{-0.132}$ & $0.990^{+1.265}_{-1.273}$ & $-0.277^{+9.680}_{-9.557}$ & $0.473^{+19.992}_{-20.496}$ \\ \hline
$[888,928]$  & 0.155 & $67.68^{+0.40}_{-0.39}$ & $-0.535^{+0.125}_{-0.131}$ & $0.975^{+1.289}_{-1.269}$ & $-0.137^{+9.647}_{-9.727}$ & $0.084^{+20.201}_{-20.269}$ \\ \hline
$[928,968]$  & 0.149 & $67.61^{+0.38}_{-0.37}$ & $-0.523^{+0.122}_{-0.127}$ & $1.017^{+1.274}_{-1.288}$ & $0.494^{+9.944}_{-9.658}$ & $0.179^{+20.107}_{-20.380}$ \\ \hline
$[968,1000]$ & 0.144 & $67.68^{+0.38}_{-0.37}$ & $-0.536^{+0.121}_{-0.127}$ & $1.022^{+1.291}_{-1.312}$ & $-0.514^{+10.041}_{-9.708}$ & $0.397^{+20.059}_{-20.379}$ \\ \hline

\end{tabular}
\end{adjustbox}

\caption{Cosmographic parameter estimates for DECIGO as a function of the signal-to-noise ratio. The quantity $\bar{z}$ denotes the mean redshift associated with each SNR interval.}
\label{tab:DECIGO_SNR}
\end{table}









\end{document}